\documentclass[a4paper,11pt]{article}
\pdfoutput=1

\usepackage{jheppub} 
\usepackage[T1]{fontenc}
\usepackage[utf8]{inputenc}
\usepackage[english]{babel}
\usepackage{amsmath,bbm}
\usepackage{amsfonts}
\usepackage{amssymb}
\usepackage{graphicx}
\usepackage[normalem]{ulem}

\usepackage[usenames,dvipsnames]{xcolor}
\usepackage[compat=1.1.0]{tikz-feynman}
\usepackage{hyperref}

\hypersetup{unicode,psdextra}

\newcommand*{\smgroup}{\mbox{$\mathit{SU}(3)_C \times \mathit{SU}(2)_L \times U(1)_Y$}}
\newcommand*{\eweakgroup}{\mbox{$\mathit{SU}(2)_L \times U(1)_Y$}}
\newcommand*{\emgroup}{\mbox{$U(1)_{\mathrm{em}}$}}

\newcommand*{\matheweakgroup}{\mathit{SU}(2)_L \times U(1)_Y}
\newcommand*{\mathemgroup}{U(1)_{em}}

\newcommand*{\unitmatrix}{\mathbbm{1}}
\newcommand*{\twomat}[1]{\underline{#1}}             
\newcommand*{\tvec}[1]{\boldsymbol{#1}}              
\newcommand*{\im}[1]{\text{Im} {#1}}                        
\newcommand*{\re}[1]{\text{Re} {#1}}                        

\newcommand*{\trans}{\mathrm{T}}                     

\newcommand{\KT}{\widetilde{\tvec{K}}}

\newcommand{\gT}{\widetilde{g}}
\newcommand{\gK}{\big( \gT \KT \big)}

\DeclareMathOperator{\diag}{diag}		
\DeclareMathOperator{\tr}{Tr}		

\title{The general Two-Higgs Doublet Model in a gauge-invariant form}

\author[a]{L.~Sartore}
\author[b]{M.~Maniatis}
\author[a]{I.~Schienbein}
\author[c]{B.~Herrmann}

\affiliation[a]{Laboratoire de Physique Subatomique et de Cosmologie, Universit\'e Grenoble-Alpes, CNRS/IN2P3, 53 Avenue des Martyrs, 38026 Grenoble, France}
\affiliation[b]{Centro de Ciencias Exactas,  Universidad del B\'io-B\'io,  Casilla  447,  Chill\'{a}n, Chile}
\affiliation[c]{LAPTh, Univ.\ Savoie Mont Blanc, CNRS, F-74000 Annecy, France}

\emailAdd{lohan.sartore@lpsc.in2p3.fr}
\emailAdd{maniatis8@gmail.com}
\emailAdd{ingo.schienbein@lpsc.in2p3.fr}
\emailAdd{herrmann@lapth.cnrs.fr}

\preprint{LAPTH-045/22}

\abstract{In the general Two-Higgs Doublet Model it has been shown that the Higgs potential can be expressed in terms of  gauge-independent quantities. In particular, stability, electroweak symmetry breaking, and CP symmetry can be understood in a concise way, avoiding unphysical gauge degrees of freedom. We complete this program and show how all the masses, the trilinear and quartic scalar interactions, the gauge-boson-Higgs interactions, as well as the Yukawa couplings in the general THDM can be expressed in a gauge-invariant way.
}
\begin{document} 
\maketitle
\flushbottom


\section{Introduction}
\label{sec:introduction}

The Two-Higgs Doublet Model (THDM) was introduced by T.~D.~Lee decades ago~\cite{Lee:1973iz} in order to have another source for CP violation -- necessary in particular to address 
the imbalance problem of baryonic matter over baryonic antimatter in the observable Universe~\cite{Sakharov:1967dj}. Still today this argument remains one of the main motivations to look for physics beyond the Standard Model of elementary particles~\cite{Zyla:2020zbs}. 

The electroweak precision measurements give strong restrictions for extensions
of the Standard Model with respect to further Higgs multiplets. In particular the
$\rho$ parameter~\cite{Zyla:2020zbs}, relating the electroweak gauge boson masses with the weak mixing angle, is in very good agreement with the prediction of the Standard Model.
In general, the $\rho$ parameter is very sensitive to any new Higgs multiplets~\cite{Bernreuther:1998rx} which couple to the electroweak gauge bosons. 
However, at least at tree level, the $\rho$ parameter remains unchanged for an arbitrary number of Higgs-boson doublets in the model. This opens the possibility to study models with an extended number of Higgs-boson doublets. Here we want to consider the simplest extension with two Higgs-boson doublets, the Two-Higgs Doublet model~(THDM).
In the past decades lots of effort has been spent since the seminal work of T.~D.\ Lee in the THDM; we refer in this respect to the review~\cite{Branco:2011iw} and some newer works like~\cite{Anisha:2022hgv,Basler:2017uxn,Athron:2022gga, Arco:2022xum,Ferreira:2020ana,Jana:2020pxx,Bednyakov:2018cmx,Cai:2018tet,Goudelis:2013uca,Belanger:2015kga}.

Extending the Standard Model to two Higgs-boson doublets,
it is clear that the model has a much richer structure: First of all, the most
general Higgs potential~\cite{Gunion:1989we} which can be formed out of two doublets contains already 14 real parameters compared to two in the Standard Model. Of course, not all parameters can be chosen arbitrarily. For instance, in order to achieve stability, that is, a potential bounded from below, the parameters are restricted.
Similar, in order to achieve the observed electroweak-symmetry breaking \eweakgroup $\to$ \emgroup, further restrictions appear. In addition, following the philosophy of the Standard Model to write down all Lagrangian terms not violating any symmetry of the model and not violating renormalisability, the most general Yukawa couplings of the two-Higgs boson doublets with the fermions of the model yield large flavour-changing neutral currents -- not observed in Nature. The usual way to avoid these unobserved interactions is to impose symmetries on the model. A simple example is a $\mathbbm{Z}_2$ symmetry, such that only one Higgs-boson doublet transforms non-trivially with a sign flip. This restricts the Yukawa couplings to only one of the two Higgs-boson doublets. Symmetries of the general THDM have been studied in some detail; see for instance~\cite{Ivanov:2007de, Maniatis:2011qu, Ferreira:2020ana, Bento:2020jei}.

It has been shown that stability and electroweak symmetry breaking of any THDM can be studied concisely with the help of {\em bilinears}~\cite{Nagel:2004sw, Maniatis:2006fs, Nishi:2006tg}. In particular, all unphysical gauge degrees of freedom of the Higgs doublets are systematically avoided. 
The principal idea is to consider the gauge-invariant scalar products of the doublets instead of the gauge-dependent doublet components themselves. This formalism has turned out to be rather powerful not only for the study of stability and the electroweak symmetry breaking but also to study symmetries of the potential or the renormalisation-group equations~\cite{Maniatis:2020wfz, Sartore:2020gou}. 
Since the gauge dependency obscures the physical content, new insights could be gained based on the gauge-invariant formalism.

In this work we want to show how to extend this formalism
to the complete model, avoiding systematically gauge redundancies in the 
scalar trilinear and quartic couplings,
the gauge-boson-Higgs interactions, and the
Yukawa couplings. We shall illustrate in examples how this can be achieved together with the advantages of the method.

In the scalar sector, the starting point is to derive the squared 
mass matrix in a gauge-invariant way. We shall see that this can be done without resorting to the actual form of the potential. Therefore, the resulting expressions are valid for any THDM potential, at tree-level or effective, to any perturbation order. 
In turn, the trilinear and quartic couplings are obtained by taking suitable derivatives of the mass matrix, leading to concise and gauge-invariant expressions which reveal new insights. For instance, we find an all-order expression for the mass of the charged Higgs boson, valid for any THDM.

Let us briefly outline the paper:
In order to make this work self-contained, we briefly recall in section~\ref{sec:bilinear} the bilinear formalism and show how the electroweak-symmetry breaking behaviour can be studied concisely.
In section~\ref{sec:gaugeInvariantFormalism}, we derive the scalar mass matrix to all orders in perturbation theory in the THDM
and apply these results to a general tree-level THDM potential. 
Then, we derive gauge invariant expressions for the trilinear and quartic scalar interactions, the gauge-boson-Higgs interactions, and the Yukawa interactions in the respective sections~\ref{sec:treeLeveLScalarCouplings}--\ref{sec:yukawa}.
Finally, in section~\ref{sec:conclusions}, we summarise our main results and draw some conclusions.
A more detailed discussion of the structure of the scalar mass matrix, extending section~\ref{sec:gaugeInvariantFormalism},
is presented in appendix~\ref{app:scalarMassMatrixStructure}.
A list of analytic expressions for all relevant tree-level couplings of the general THDM in gauge-invariant form
can be found in appendix~\ref{app:couplings}.


\section{Review of the bilinear formalism}
\label{sec:bilinear}

Here, we briefly review the bilinears in the THDM~\cite{Nagel:2004sw, Nishi:2006tg, Maniatis:2006fs} in order to make this article 
self contained. We will also discuss basis transformations. In the convention with both Higgs-boson doublets carrying hypercharge $y=+1/2$ corresponding to upper charged components, we write
\begin{equation} \label{eq:doublets}
\begin{split}
&\varphi_1(x) = \begin{pmatrix} \varphi_1^{+}(x)\\ \varphi_1^{0}(x) \end{pmatrix} =
\frac{1}{\sqrt{2}} 
\begin{pmatrix} \pi_1^1(x)+i \sigma_1^1(x)\\ \pi_1^2(x)+i \sigma_1^2(x) \end{pmatrix}
,\\
&\varphi_2(x) = \begin{pmatrix} \varphi_2^{+}(x)\\ \varphi_2^{0}(x) \end{pmatrix} =
\frac{1}{\sqrt{2}} 
\begin{pmatrix} \pi_2^1(x)+i \sigma_2^1(x)\\ \pi_2^2(x)+i \sigma_2^2(x) \end{pmatrix}.
\end{split}
\end{equation}
Explicitly we have decomposed the complex fields into their real and imaginary components.
From now on we suppress the space-time argument of the fields. We will also write the real and imaginary upper and lower components of the two doublets in the form of one eight-component vector:
\begin{equation} \label{eq:phi}
    \phi = \begin{pmatrix}
    \pi_1^1, \pi_1^2, \sigma_1^1, \sigma_1^2, \pi_2^1, \pi_2^2, \sigma_2^1, \sigma_2^2\end{pmatrix}^\trans
    , \quad \text{with }
    \phi \in \mathbb{R}^8\,. 
\end{equation}

The most general, tree-level, gauge-invariant potential with two Higgs-boson doublets reads~\cite{Gunion:1989we}
\begin{equation}
\label{eq:Vconv}
\begin{split}
V^0_{\text{THDM}} (\varphi_1, \varphi_2) =~& 
m_{11}^2 (\varphi_1^\dagger \varphi_1) +
m_{22}^2 (\varphi_2^\dagger \varphi_2) -
m_{12}^2 (\varphi_1^\dagger \varphi_2) -
(m_{12}^2)^* (\varphi_2^\dagger \varphi_1)\\
& +\frac{1}{2} \lambda_1 (\varphi_1^\dagger \varphi_1)^2 
+ \frac{1}{2} \lambda_2 (\varphi_2^\dagger \varphi_2)^2 
+ \lambda_3 (\varphi_1^\dagger \varphi_1)(\varphi_2^\dagger \varphi_2) \\ 
&+ \lambda_4 (\varphi_1^\dagger \varphi_2)(\varphi_2^\dagger \varphi_1)
+ \frac{1}{2} \Big[ \lambda_5 (\varphi_1^\dagger \varphi_2)^2 + \lambda_5^* 
(\varphi_2^\dagger \varphi_1)^2 \Big] \\ 
&+ \Big[ \lambda_6 (\varphi_1^\dagger \varphi_2) + \lambda_6^* 
(\varphi_2^\dagger \varphi_1) \Big] (\varphi_1^\dagger \varphi_1) + \Big[ \lambda_7 (\varphi_1^\dagger 
\varphi_2) + \lambda_7^* (\varphi_2^\dagger \varphi_1) \Big] (\varphi_2^\dagger \varphi_2).
\end{split}
\end{equation}
In this potential we have two real quadratic parameters~$m_{11}^2$, $m_{22}^2$, and one complex quadratic parameter~$m_{12}^2$ as well as seven quartic parameters $\lambda_1, \dots, \lambda_7$, where the first four are real and $\lambda_5$, $\lambda_6$, $\lambda_7$ complex. Altogether this corresponds to 14~real parameters in the potential. 

Bilinears systematically avoid unphysical gauge degrees of freedom and are defined in the following way: 
First, the $2 \times 2$ matrix of the two doublets,
\begin{equation} \label{eq:defphi}
    \psi = \begin{pmatrix} \varphi_1^\trans \\ \varphi_2^\trans \end{pmatrix} 
    = \begin{pmatrix} \varphi_1^+& \varphi_1^0 \\ \varphi_2^+& \varphi_2^0 \end{pmatrix} \, ,
\end{equation}
is introduced. 
All gauge-invariant scalar products of the two doublets $\varphi_1$ and $\varphi_2$  can be arranged in one matrix 
\begin{equation} \label{eq:Kbar}
\twomat{K} = {\psi} {\psi}^\dagger =
\begin{pmatrix}
 \varphi_1^\dagger \varphi_1 &  ~\varphi_2^\dagger \varphi_1\\
 \varphi_1^\dagger \varphi_2 &  ~\varphi_2^\dagger \varphi_2
 \end{pmatrix}.
\end{equation}

This  by construction hermitian $2 \times 2$ matrix~$\twomat{K}$ can be decomposed into a basis of the unit matrix and the Pauli matrices,
\begin{equation}
 \twomat{K} = \frac{1}{2} \bigg( K_0 \unitmatrix_2 + K_a \sigma_a \bigg), \qquad a=1,2,3, \label{eq:KbarToBilinears}
\end{equation}
with the usual convention to sum over repeated indices. The four real coefficients $K_0$, $K_a$ are called bilinears. Building traces on both sides of this equation (also with products of Pauli matrices) we get the four real bilinears explicitly,
\begin{align}
\label{eq:Kphi}
&K_0 = \varphi_1^\dagger \varphi_1 + \varphi_2^\dagger \varphi_2,
&&K_1 = \varphi_1^\dagger \varphi_2 + \varphi_2^\dagger \varphi_1, \nonumber \\
&K_2 = i\big( \varphi_2^\dagger \varphi_1 - \varphi_1^\dagger \varphi_2 \big), 
&&K_3 = \varphi_1^\dagger \varphi_1 - \varphi_2^\dagger \varphi_2. 
\end{align}
Inverting these relations we can express any THDM potential in terms of bilinears:
\begin{align}
\label{eq:phiK}  
& \varphi_1^\dagger \varphi_1 = \frac{1}{2}\left( K_0 + K_3\right),
&& \varphi_1^\dagger \varphi_2 = \frac{1}{2}\left( K_1 + i K_2\right), \nonumber\\
& \varphi_2^\dagger \varphi_1 = \frac{1}{2}\left( K_1 - i K_2\right),
&& \varphi_2^\dagger \varphi_2 = \frac{1}{2}\left( K_0 - K_3\right).
\end{align}
The matrix $\twomat{K}$ is positive semi-definite.
From $K_0 = \tr (\twomat{K})$  and $\det(\twomat{K}) = \tfrac{1}{4} (K_0^2 -K_a K_a)$
we obtain 
\begin{equation} \label{eq:Kdom}
K_0 \ge 0, \qquad K_0^2 - K_a K_a \ge 0.
\end{equation}
As has been shown in~\cite{Maniatis:2006fs} there is a one-to-one correspondence between the original doublet fields and the bilinears apart from unphysical gauge-degrees of freedom. In terms of bilinears we can write any tree-level THDM potential (a constant term can always be dropped) as
\begin{equation}
\label{eq:pot}
V^0_{\text{THDM}} (K_0, K_a)=
 \xi_0 K_0 + \xi_a K_a + \eta_{00} K_0^2 + 2 K_0 \eta_a K_a + K_a E_{ab} K_b,
\end{equation}
with real parameters $\xi_0$, $\xi_a$, $\eta_{00}$, $\eta_a$, $E_{ab} = E_{ba}$, $a,b \in \{1,2,3\}$.
Expressed in terms of the conventional parameters appearing in~\eqref{eq:Vconv}, these new parameters read
\begin{align}
\xi_0 &= \frac{1}{2}\left(m_{11}^2+m_{22}^2\right) ,
\quad
\tvec{\xi} = (\xi_\alpha)=\frac{1}{2}
\begin{pmatrix}
- 2 \re(m_{12}^2), &
 2 \im(m_{12}^2), &
 m_{11}^2-m_{22}^2
\end{pmatrix}^\trans,
\label{eq:para1}
\\
\eta_{00} &= \frac{1}{8}
(\lambda_1 + \lambda_2) + \frac{1}{4} \lambda_3  ,
\quad
\tvec{\eta} = (\eta_{a})=\frac{1}{4}
\begin{pmatrix}
\re(\lambda_6+\lambda_7), & 
-\im(\lambda_6+\lambda_7), & 
\frac{1}{2}(\lambda_1 - \lambda_2)
\end{pmatrix}^\trans, 
\label{eq:para2}
\\
E &= (E_{ab})= \frac{1}{4}
\begin{pmatrix}
\lambda_4 + \re(\lambda_5) & 
-\im(\lambda_5) & \re(\lambda_6-\lambda_7) 
\\ 
-\im(\lambda_5) & \lambda_4 - \re(\lambda_5) & 
-\im(\lambda_6-\lambda_7) \\ 
\re(\lambda_6-\lambda_7) & 
-\im(\lambda_6 -\lambda_7) & 
\frac{1}{2}(\lambda_1 + \lambda_2) - \lambda_3
\end{pmatrix}.
\label{eq:para3}
\end{align}

As has been shown in~\cite{Maniatis:2006fs}, we can also form Minkowski-type four-vectors 
from the bilinears,
\begin{equation}
    \KT=\begin{pmatrix} K_0\\ \tvec{K}\end{pmatrix}, \quad
    \text{with } \tvec{K}=
    \begin{pmatrix} K_1\\ K_2\\ K_3 \end{pmatrix}.
\end{equation}
Writing the parameters of the potential in the form
\begin{equation}
    \tilde{\tvec{\xi}}=\begin{pmatrix} \xi_0 \\ \tvec{\xi}\end{pmatrix},
    \qquad
    \tilde{E}=\begin{pmatrix} \eta_{00} & \tvec{\eta}^\trans\\
    \tvec{\eta} & E \end{pmatrix},
\end{equation}
the tree-level potential~\eqref{eq:pot} can be written as
\begin{equation} \label{eq:potK4}
    V^0_{\text{THDM}}(\KT) =  \KT^\trans \tilde{\tvec{\xi}} + 
     \KT^\trans \tilde{E}  \KT.
\end{equation}
The conditions~\eqref{eq:Kdom} for the bilinears can, with the help of
\begin{equation}
    \tilde{g} = \diag(1, -\unitmatrix_3)\,,
\end{equation}
be expressed as
\begin{equation} \label{eq:dom4}
    K_0 \ge 0, \qquad \KT^\trans \tilde{g} \KT \ge 0\;.
\end{equation}

We will consider unitary mixing of the two doublets,
\begin{equation} \label{eq:basistf}
\varphi'_i = U_{ij} \varphi_j, \quad \text{with } U = (U_{ij}),  \quad U^\dagger U = \unitmatrix_2\,.
\end{equation}
These basis transformations correspond in terms of bilinears to
\begin{equation} \label{eq:b1}
K_0' = K_0, \quad K_a' = R_{ab}(U) K_b \;,
\end{equation}
with $R_{ab}(U)$ defined by
\begin{equation} \label{eq:bilinearRtransfo}
U^\dagger \sigma^a U = R_{ab}(U) \sigma^b \;.
\end{equation}
It follows that $R(U) \in SO(3)$, that is, $R(U)$ is a proper rotation in three dimensions. 
We see that the potential~\eqref{eq:pot} stays invariant under a 
change of basis of the bilinears \eqref{eq:b1}
if we simultaneously transform the parameters~\cite{Maniatis:2006fs}
\begin{equation} \label{eq:b2}
\xi'_0 = \xi_0, \quad
\xi_a' = R_{ab} \xi_b, \quad
\eta_{00}' = \eta_{00}, \quad
\eta_{a}' = R_{ab} \eta_{b}, \quad
E_{cd}' = R_{ca}  E_{ab} R_{bd}^\trans \, .
\end{equation}
Note that by a change of basis we 
can diagonalise the real symmetric matrix~$E$. 

Let us also briefly recall from~\cite{Maniatis:2006fs} how electroweak symmetry breaking translates into conditions on the gauge-invariant bilinears. First, we suppose that the potential of the THDM is stable, that is, it is bounded from below. Then at a minimum of the potential~\eqref{eq:defphi} becomes, 
\begin{equation} \label{eq:phivev}
    \langle \psi \rangle 
    = \begin{pmatrix} v_1^+& v_1^0 \\ v_2^+& v_2^0 \end{pmatrix}.
\end{equation}
Here, we have introduced the vacuum expectation values of the components of the doublets, that is, 
$v^+_{1/2} = \langle \varphi^+_{1/2} \rangle$ and
$v^0_{1/2} = \langle \varphi^0_{1/2} \rangle$. 

In case that we have a charge-breaking (CB) minimum, the matrix $\langle \psi \rangle$, and therefore also $\langle \twomat{K} \rangle = \langle \psi \rangle \langle \psi \rangle^\dagger$, has full rank and this translates to the condition (suppressing the angular brackets in the notation indicating the vacuum)
\begin{equation} \label{bEW}
\text{CB:} \qquad K_0>0, \qquad K_0^2 -\tvec{K}^\trans \tvec{K}  = \KT^\trans \tilde{g} \KT > 0\;.
\end{equation}

In contrast, in case we have a neutral, charge-conserving (CC) vacuum, the matrix $\langle \psi \rangle$ together with $\langle \twomat{K} \rangle$ has to have rank one resulting in
\begin{equation} \label{EW}
\text{CC:} \qquad K_0>0, \qquad K_0^2 -\tvec{K}^\trans \tvec{K}  = \KT^\trans \tilde{g} \KT = 0\;.
\end{equation}
A vacuum which does not break the electroweak symmetry at all corresponds to
\begin{equation}
    K_0 = 0\,,
\end{equation}
where we also have $\KT=0$.

In case of a charge-conserving minimum we can, by a basis change, achieve that only the component~$\varphi_1^0$ gets a non-vanishing vacuum-expectation value and we set in this basis
\begin{equation} \label{eq:v10beta}
    \langle {\varphi_1^0}'\rangle = 
    {v_1^0}' \equiv \frac{v}{\sqrt{2}} 
\end{equation}
with the conventional factor $1/\sqrt{2}$. This basis, in which only the neutral component of $\varphi_1$ gets a non-vanishing vacuum-expectation value, is sometimes called Higgs basis. The bilinears in this basis read $\KT=\begin{pmatrix} v^2/2,& 0, & 0,& v^2/2 \end{pmatrix}^\trans$.

For a stable potential, the minimum can be found from the gradient of the potential. In the case of a charge-breaking minimum, the conditions read
\begin{equation} \label{eqbEW}
\text{CB:} \qquad 
\partial_\mu V \equiv \frac{\partial V}{\partial K^\mu} = 0
\end{equation}
on the domain~\eqref{bEW}. An unbroken minimum is simply given 
for vanishing doublets, corresponding to $K_0=0$ and therefore for 
a vanishing potential.
The correct electroweak-symmetry breaking minimum can be found by introducing a Lagrange multiplier~$u$ in order to satisfy the second equation in~\eqref{EW}, that is, for a minimum with the correct electroweak symmetry breaking, 
\begin{equation} \label{eqEW}
\text{CC:} \qquad 
\partial_\mu V =2 u  (\tilde{g} \KT)_\mu
\end{equation}
on the domain~\eqref{EW}.

Let us also recall the dimensionless expressions of the bilinears. First we note that for $K_0=0$ the potential is trivially vanishing.
We define for $K_0 > 0$~\cite{Maniatis:2006fs}
\begin{equation}
\label{eq-ksde}
k_a = \frac{K_a}{K_0}, \quad a \in \{1,2,3\} \quad \text{and} \quad 
\tvec{k} = 
\begin{pmatrix} k_1, & k_2, & k_3 \end{pmatrix}^\trans \;.
\end{equation}
With \eqref{eq-ksde}
we can write the tree-level potential~\eqref{eq:pot}
in the form
\begin{equation}
\label{eq-vk}
V^0_{\text{THDM}} = K_0
\left(\xi_0 + \tvec{\xi}^\trans \tvec{k}\right)
+ K_0^2 \left(\eta_{00} 
  + 2 \tvec{\eta}^\trans \tvec{k} + \tvec{k}^\trans E \tvec{k} \right)
\end{equation}
defined on the compact domain, as follows from~\eqref{eq:Kdom},
\begin{equation} \label{domk}
|\tvec{k}| \leq 1 \;.
\end{equation}

We will follow the convention to use Greek indices $\mu$, $\nu,  \ldots \in \{0, \ldots, 3\}$ for the Minkowski-type four-vectors, for instance we write $K_\mu$. The two doublets themselves are distinguished by Latin indices $i$, $j, \ldots \in \{1,2\}$. For the component fields 
of the two doublets~\eqref{eq:phi} we also use Latin indices $i$, $j, \ldots \in \{1, \ldots, 8\}$. Let us note that we do not distinguish between upper and lower indices, that is, for instance we have $K^\mu = K_\mu$ with  $(K^\mu)=(K_\mu)= \begin{pmatrix} K_0, K_1, K_2, K_3 \end{pmatrix}^\trans$ as well as for the 8-component vector $\phi^i = \phi_i$.


\section{Gauge-invariant scalar mass matrices}
\label{sec:gaugeInvariantFormalism}

The conventional procedure to compute the scalar mass matrices in the THDM is to firstly check that the potential is stable, that is, bounded from below. Secondly, a minimum (typically the global minimum) has to be found. Thirdly, the electroweak symmetry-breaking behaviour of the minimum has to be checked. For a physically acceptable minimum it has to be verified that the electroweak symmetry is broken 
$\eweakgroup \to \emgroup$. At a fixed gauge, the second derivatives of the potential with respect to the excitation fields about the vacuum give then the squared mass matrices.

In contrast, here we want to express the mass matrices in terms of gauge-invariant quantities. The mass matrices are defined as the second derivative of the Lagrangian with respect to the fields. Therefore we have first to establish the connection of the component fields of the doublets~\eqref{eq:phi}, $\phi^i$, $i \in \{1, \ldots, 8\}$ 
to the bilinear fields $K^\mu$, $\mu \in \{0, \ldots 3\}$.

The gauge invariants, that is, the bilinear fields can be written in terms of the components of the doublets~\eqref{eq:phi}
\begin{equation}\label{eq:KphiDef}
	K^\mu \equiv \frac{1}{2} \Delta^\mu_{ij} \phi^i \phi^j\,,
	\qquad i,j \in \{1,\ldots,8\}.
\end{equation}
For instance, in the basis given in~\eqref{eq:phi}, we find explicitly the four real and symmetric $8 \times 8$ matrices $\Delta^\mu_{ij}$:
\begin{align}
\label{eq:DeltaMatrices}
	&\Delta^0 = \begin{pmatrix}
		\phantom{+}\unitmatrix_2 & & & \\
		& \phantom{+}\unitmatrix_2 & & \\
		& &\phantom{+}\unitmatrix_2 &  \\
		& & &\phantom{+}\unitmatrix_2
	\end{pmatrix}, \qquad
	&&\Delta^1 = \begin{pmatrix}
		& & \phantom{+}\unitmatrix_2 & \\
		& & & \phantom{+}\unitmatrix_2 \\
		\phantom{+}\unitmatrix_2 & & &  \\
		& \phantom{+}\unitmatrix_2 & &
	\end{pmatrix}, \nonumber\\
	&\Delta^2 = \begin{pmatrix}
		& & & \phantom{+}\unitmatrix_2 \\
		& & -\unitmatrix_2 & \\
		& -\unitmatrix_2 & &  \\
		\phantom{+}\unitmatrix_2 & & &
	\end{pmatrix}, \qquad
	&&\Delta^3 = \begin{pmatrix}
		\phantom{+}\unitmatrix_2 & & & \\
		& \phantom{+}\unitmatrix_2 & & \\
		& &-\unitmatrix_2 &  \\
		& & &-\unitmatrix_2
	\end{pmatrix}\, ,
\end{align}
where $\unitmatrix_2$ is the $2 \times 2$ unit matrix and the the blank spaces have to be filled with zeros.
It should be noted that the matrices $\Delta^{1,2,3}$ resemble the Pauli matrices $\sigma^{1,2,3}$, respectively, in that
they satisfy a Clifford algebra $\{\Delta^a,\Delta^b\}= 2 \delta^{ab} \unitmatrix_8$ (but not a Lie algebra).

The connection between the bilinears and the component fields are therefore given by the four $8 \times 4$ matrices $\Gamma$, defined as
\begin{equation} \label{eq:defgamma}
   \Gamma^\mu_i \equiv \frac{\partial K^\mu}{\partial \phi^i} = \partial_i K^\mu = \Delta^\mu_{ij} \phi^j  \,.
\end{equation}
Note that the Greek indices correspond to 
the gauge invariants, whereas the Latin indices correspond to the component fields $\phi^i$.

We would now like to derive the algebra of the
four $8 \times 8$ matrices $\Delta_{ij}^\mu$. To this
end we compute
\begin{align}\label{eq:ga21}
    \left(\Gamma^2\right)^{\mu\nu} = \left(\Gamma^\trans \Gamma\right)^{\mu\nu} = \Gamma^\mu_i \Gamma^\nu_i = \Delta^\mu_{ij} \Delta^\nu_{ik}\, \phi^j \phi^k  
    = \frac{1}{2}\left\{\Delta^\mu, \Delta^\nu\right\}_{jk} \phi^j \phi^k\, ,
\end{align}
where $\{\Delta^\mu, \Delta^\nu\}$ is the anti-commutator of the $\Delta$-matrices.
 From~\eqref{eq:ga21}, we observe that $\Gamma^2$ is a gauge-independent object (since it only carries Greek bilinear-field indices) and that it depends quadratically on the fields. Therefore, a rank-3 tensor $T_\lambda^{\mu \nu}$ can be defined such that
\begin{equation}\label{eq:ga22}
    \left(\Gamma^2\right)^{\mu\nu} = T_\lambda^{\mu\nu} K^\lambda\,.
\end{equation}
It then follows from Eqs.~\eqref{eq:KphiDef}, \eqref{eq:ga21} and \eqref{eq:ga22} that
\begin{equation}\label{eq:anti-comm}
    \left\{\Delta^\mu, \Delta^\nu\right\}_{ij} =  T^{\mu\nu}_\lambda \Delta^\lambda_{ij}\,.
\end{equation}
Observing that the symmetric $\Delta$-matrices obey a closure relation
\begin{equation}
    \Delta^\mu_{ij} \Delta^\nu_{ij} = \Delta^\mu_{ij} \Delta^\nu_{ji} =\tr (\Delta^\mu \Delta^\nu) = 8 \delta^{\mu\nu}\,,
\end{equation}
we can provide an explicit expression\footnote{Note that an alternative expression can be given in terms of Pauli matrices:
\begin{equation*}
   T^{\mu\nu}_\lambda = \frac{1}{2}\mathrm{Tr}\,\big(\{\sigma^\mu, \sigma^\nu\}\,\sigma_\lambda\big)\,.
\end{equation*}
} for $T^{\mu\nu}_\lambda$:
\begin{equation} \label{eq:Texp}
    T^{\mu\nu}_\lambda = \frac{1}{8}\mathrm{Tr}\left(\left\{\Delta^\mu, \Delta^\nu\right\} \Delta_\lambda\right)\,.
\end{equation}
In particular, we obtain 
\begin{equation}\label{eq:Gamma2expr}
    \Gamma^2 = 2 \begin{pmatrix}
       K_0 & K_1 & K_2 & K_3 \\ K_1 & K_0 & 0 & 0 \\ K_2 & 0 & K_0 & 0 \\ K_3 & 0 & 0 & K_0
    \end{pmatrix} = 2 \begin{pmatrix}
       K_0 & \tvec{K}^\trans\\ \tvec{K} & K_0 \unitmatrix_3
    \end{pmatrix}\,.
\end{equation}
Since $\det (\Gamma^2) = 16 K_0^2 \left(K_0^2 - \tvec{K}^2\right) = 16 K_0^2 \, \KT^\trans\gT\KT$, the electroweak symmetry-breaking behaviour depends on the nullspace structure of $\Gamma^2$. 
As discussed in the last section, a charge-breaking minimum satisfying~\eqref{bEW} leads to a complete breakdown of the $\eweakgroup$ 
group whereas \eqref{EW} leads to the viable, charge-conserving $\matheweakgroup \rightarrow \mathemgroup$ breaking. 
We now consider the scalar mass matrix with respect to the charge-breaking~(CB) and charge-conserving~(CC) case separately.

The field-dependent scalar mass matrix is expressed in our formalism as
\begin{equation} \label{eq:msExpr}
    \left(M_s^2\right)_{ij} = \partial_i \partial_j V = \partial_i \left(\Gamma^\mu_j \partial_\mu V \right) = \Delta^\mu_{ij} \partial_\mu V + \Gamma^\mu_i \Gamma^\nu_j \partial_\mu \partial_\nu V \,.
\end{equation}
With the definition
\begin{equation} \label{eq:Mmunu}
    \mathcal{M}_{\mu\nu} = \partial_\mu \partial_\nu V\,
\end{equation}
and writing $\mathcal{M} = (\mathcal{M}_{\mu\nu})$
as well as $\Delta^\mu =(\Delta^\mu_{ij})$ 
we express~\eqref{eq:msExpr} in matrix form,
\begin{equation} \label{eq:msExprMat}
    M_s^2 = \Delta^\mu \partial_\mu V + \Gamma \mathcal{M} \Gamma^\trans.
\end{equation}
With this form we have achieved an intuitive understanding of the scalar spectrum at a minimum of the potential. In particular, for a charge-breaking minimum, where $\partial_\mu V = 0$ according to~\eqref{eqbEW}, the $8 \times 8$ scalar mass matrix reduces to
\begin{equation}\label{eq:MsCB}
    M_s^2 \stackrel{CB}{=} \Gamma \mathcal{M} \Gamma^\trans
\end{equation}
and can only possess four non-zero eigenvalues (\textit{i.e.}~as many as the number of independent gauge-invariant scalar operators). The four remaining massless states correspond to the would-be Goldstone bosons associated with the four broken generators of the {\eweakgroup}
gauge group. On the other hand, as we will show below, at a charge-conserving minimum, where $\partial_\mu V =2 u  (\tilde{g} \KT)_\mu$ or $\det(\Gamma^2) = 0$ according to \eqref{eqEW}, the $\Gamma \mathcal{M} \Gamma^\trans$ matrix
possesses at most three non-zero eigenvalues and the full scalar mass matrix
\begin{equation}\label{eq:MsCC}
    M_s^2 \stackrel{CC}{=} 2u \big(\gT \KT\big)_\mu \Delta^\mu + \Gamma \mathcal{M} \Gamma^\trans
\end{equation}
contains five massive states and three Goldstone modes corresponding to the broken generators of $\matheweakgroup\rightarrow\mathemgroup$. 
Now we want to consider an orthogonal rotation of
the component fields,
\begin{equation}
    \widehat{\phi}^i = U^{ij} \phi^j \quad 
    \text{with} \quad U^{ij} U^{kj} = \delta^{ik}, \quad
    i,j,k \in \{1,\ldots,8\}\,.
\end{equation}
In matrix notation we write simply
$\widehat{\phi}= U \phi$ with $U^\trans U = \unitmatrix_8$.
By an orthogonal rotation - only applied to the component fields, leaving the gauge invariants $\KT$ unchanged - we 
can always achieve a form for the 
$8\times4$ matrix $\Gamma$ with
the first four rows vanishing\footnote{The matrix $\Gamma$ consists of four 8-component vectors spanning a
4-dimensional hypersurface. It is always possible to choose a basis such that the 8-4 dimensional space orthogonal 
to the hypersurface is represented by the first four basis vectors such that the corresponding components of the four 8-component vectors are zero.},
\begin{equation} \label{eq:canonicalGamma}
    \widehat{\Gamma} = U_c\, \Gamma = \begin{pmatrix}
    0_{4\times4} \\ \gamma
    \end{pmatrix}\,,
\end{equation}
where we will call~$U_c$
a canonical rotation and the 
corresponding bases, 
where $\widehat{\Gamma}$ has this form, canonical bases. 
The matrix $\gamma$ is obviously a $4\times4$ matrix. 
We get
\begin{equation}
    \Gamma^2 = \Gamma^\trans \Gamma = \widehat{\Gamma}^\trans \widehat{\Gamma} = 
    \gamma^\trans \gamma
    \,,
\end{equation}
meaning that the $\gamma$ matrix can be obtained from a Cholesky-like\footnote{The usual Cholesky decomposition would have made $\gamma$ an upper-triangular matrix, while it is in our case lower-triangular.} decomposition of $\Gamma^2$, and that the resulting expression only depends on gauge invariants, \textit{i.e.}~the bilinear fields. Requiring $\gamma$ to be lower-triangular fixes its components uniquely, and we find
\begin{equation} \label{eq:gamma}
    \gamma = \sqrt{2 K_0} \begin{pmatrix}
       \sqrt{1 - \tvec{k}^\trans} \tvec{k} & \tvec{0}^\trans \\
       \tvec{k} & \unitmatrix_3
    \end{pmatrix}\,.
\end{equation}
In a canonical basis we can always express $\widehat{\Gamma}$ in terms of the field components $\widehat{\phi}=U_c \phi$:
\begin{equation}
    \widehat{\Gamma} = U_c \Gamma = U_c \Delta \phi = U_c \Delta U_c^\trans \widehat{\phi} = 
    \widehat{\Delta} \widehat{\phi}\, ,
\end{equation}
where $\widehat{\Delta}= U_c \Delta U_c^\trans$ is simply obtained from the canonical orthogonal rotation $U_c$.

Moreover, two canonical bases can always be related by an orthogonal transformation since
\begin{equation}
    \begin{pmatrix}
        U_{4\times4} & 0_{4\times4}\\
        0_{4\times4} & \unitmatrix_4
    \end{pmatrix} \widehat{\Gamma}
    = \widehat{\Gamma}\,.
\end{equation}
We want to show now that in a canonical basis the electroweak-symmetry breaking behaviour of the scalar mass matrix at the vacuum becomes manifest. First, at a charge-breaking minimum, Eq.~\eqref{eq:MsCB} directly yields\footnote{Here and in the remainder of the section, hatted quantities are understood to be expressed in a canonical basis.}
\begin{equation} \label{eq:M2sc}
    \widehat{M}_s^2 \stackrel{CB}{=} \widehat{\Gamma} \mathcal{M} \widehat{\Gamma}^\trans = \begin{pmatrix}
        0_{4\times4} & 0_{4\times4} \\
        0_{4\times4} & \gamma \mathcal{M} \gamma^\trans
    \end{pmatrix}\,.
\end{equation}
Clearly, the mass matrix expressed in this basis takes a block-diagonal form, where the Goldstone and physical sectors have been manifestly disentangled.
In addition, the number of massive states, the number of gauge singlets and the number of independent vacuum-expectation values all equal the number of independent gauge invariant operators, namely four. This equality is characteristic of a maximal breakdown of the gauge group\footnote{Generally speaking, \textit{maximally broken} does not mean that the residual gauge group is trivial, but instead that the number of gauge singlets is maximal given a specific field content. In the present case however, \eweakgroup~is indeed broken down to the trivial group.}. 

Turning to the case of a charge-conserving breaking, we show in App.~\ref{app:scalarMassMatrixStructure} 
 that the mass matrix~\eqref{eq:MsCC} in a canonical 
basis takes a block-diagonal form,
\begin{equation} \label{eq:MsqCC}
\begin{split}
    \widehat{M}_s^2  \stackrel{CC}{=} & 2u  \, \big(\gT \KT\big)_\mu \widehat{\Delta}^\mu + \widehat{\Gamma} \mathcal{M} \widehat{\Gamma}^\trans 
    \\  \stackrel{\phantom{CC}}{=} &
    2u  \begin{pmatrix}
        A_{55} & 0_{5\times3} \\
        0_{3\times5} & B_{33} \\
    \end{pmatrix}
    +
    \begin{pmatrix}
        0_{5\times5} & 0_{5\times3} \\
        0_{3\times5} & \gamma_3 \mathcal{M} \gamma_3^\trans
    \end{pmatrix}
    =
   \begin{pmatrix}
        \widehat{\mathcal{M}}^2_\mathrm{CC} & 0_{5\times3}\\
        0_{3\times5} &  \widehat{\mathcal{M}}^2_\mathrm{neutral}
    \end{pmatrix}\,,
\end{split}
\end{equation}
where $\widehat{\mathcal{M}}^2_\mathrm{CC}$ accounts for mixed contributions between massive and Goldstone states, both possibly charged under the new residual subgroup (here, \emgroup). 
The matrix $\gamma_3$, which appears in~\eqref{eq:MsqCC} is defined to be the non-vanishing, lower $3\times4$ block of the matrix~$\gamma$ evaluated at a charge-conserving minimum where $1-\tvec{k}^2=0$:
\begin{equation} \label{eq:ga3def}
    \gamma \stackrel{CC}{=} \sqrt{2 K_0} \begin{pmatrix}
        0 & \tvec{0}^\trans\\
        \tvec{k} & \unitmatrix_3
    \end{pmatrix} \equiv \begin{pmatrix}
        0_{1\times4}\\
        \gamma_3
    \end{pmatrix}\,.
\end{equation}
The matrices $A_{55}$ and $B_{33}$ read explicitly
\begin{equation}
    A_{55} = 2 K_0 \begin{pmatrix}
        0 & \tvec{0}^\trans & 0 \\
        \tvec{0} & \tvec{k} \tvec{k}^\trans & \tvec{0}\\
        0 & \tvec{0}^\trans & 1\\
    \end{pmatrix},\qquad 
    B_{33} = 2 K_0 \left(\unitmatrix_3 - \tvec{k} \tvec{k}^\trans\right) = - \gamma_3 \gT \gamma_3^\trans.\\
\end{equation}

By a further transformation from one canonical basis to another we can completely
disentangle the electrically neutral from the electrically charged contributions,
\begin{equation}\label{eq:massMatPhysical}
    \widehat{M}_s^2 \stackrel{CC}{=} \begin{pmatrix}
        0_{3\times3} & & \\
        & \widehat{\mathcal{M}}^2_\mathrm{charged} & \\
        & & \widehat{\mathcal{M}}^2_\mathrm{neutral}
    \end{pmatrix}\,.
\end{equation}
In App.~\ref{app:scalarMassMatrixStructure} we show in detail how we arrive at the form of the scalar mass matrix~\eqref{eq:massMatPhysical}. Similarly, we show that  the charged mass matrix can be diagonalised,
\begin{equation} \label{eq:allOrderMcharged}
    \bar{\mathcal{M}}^2_\mathrm{charged} = \diag\begin{pmatrix} m_{H^\pm}^2, & m_{H^\pm}^2 \end{pmatrix} = \diag\begin{pmatrix} 4 u K_0, & 4 u K_0 \end{pmatrix}\,,
\end{equation}
and the neutral scalar mass matrix reads
\begin{equation}\label{eq:allOrderMneutral}
    \widehat{\mathcal{M}}^{2}_\mathrm{neutral} = \gamma_3 \left(\mathcal{M} - 2 u \gT\right) \gamma_3^\trans\,.
\end{equation}
Having found the scalar mass matrices for the charge-conserving minimum, we go ahead and
diagonalise the neutral part~\eqref{eq:allOrderMneutral}. 
The similarity transformation to diagonalise the real symmetric matrix 
\eqref{eq:allOrderMneutral} corresponds to 
a change of basis of the bilinears. 
Let us denote with~$R$ the corresponding rotation in the 3-dimensional bilinear field space, defined as
\begin{equation}\label{eq:diagonalBasisNeutral}
    \bar{\mathcal{M}}^2_\mathrm{neutral} = R\widehat{\mathcal{M}}^2_\mathrm{neutral}R^\trans = \diag\begin{pmatrix}m_1^2, & m_2^2, & m_3^2\end{pmatrix}.
\end{equation}
As shown in Sec.~\ref{sec:bilinear}, by this basis transformation the bilinears 
transform 
as 
\begin{equation} \label{bilbasis}
    \bar{K_0} = K_0, \quad
    \bar{\tvec{K}} = R \tvec{K}\,.
\end{equation}
With respect to
the four vectors we write also
\begin{equation}
 \overline{\KT} = \widetilde{R} \KT
    \quad \text{with }
    \widetilde{R} =
    \begin{pmatrix}
    1 & \tvec{0}^\trans\\
    \tvec{0} & R
    \end{pmatrix}, 
\end{equation}
that is, $R$ is a $3 \times 3$ matrix, whereas $\tilde{R}$ a $4 \times 4$ matrix. The 
dimensionless bilinears transform as
$\bar{\tvec{k}} = R \tvec{k}$.
Similar, we find for $\Gamma^\mu_i$
and $\Delta^\mu_{ij}$ the following
transformation behaviour under a change of basis:
\begin{equation} \label{massbGD}
    \bar{\Gamma}^\mu_i = 
    \widetilde{R}^{\mu}_{\nu} \widehat{\Gamma}^\nu_i, \quad
    \bar{\Delta}^\mu_{ij} =
    \widetilde{R}^{\mu}_{\nu} 
    \widehat{\Delta}^\nu_{ij}\, \qquad
    \text{with }(\widetilde{R}^{\mu}_{\nu})=\widetilde{R}\,.
\end{equation}

After applying this rotation in the bilinear field space, we obtain the expression of $M_s^2$ in a basis which will be referred to as the \textit{mass basis} in the following. In this basis, the scalar mass matrix finally reads
\begin{equation}\label{eq:diagonalMs}
    \bar{M}_s^2 = \diag\begin{pmatrix}
        0, & 0, & 0, & m_{H^\pm}^2, & m_{H^\pm}^2, & m_1^2, & m_2^2, & m_3^2
    \end{pmatrix}\,.
\end{equation}
We finally note that the vacuum-expectation value of the scalar multiplet $\phi$, noted $\langle\phi\rangle$, can readily be computed in the mass basis from the relation
\begin{equation}
    \bar{\Gamma}^\mu_i = \bar{\Delta}^\mu_{ij} \langle\bar{\phi}\rangle^j\,,
\end{equation}
and using the expressions of $\bar{\Gamma}$ and $\bar{\Delta}$ given in Appendix~\ref{app:scalarMassMatrixStructure}. Explicitly, one has
\begin{equation} \label{eq:phiVacuum}
    \langle \bar{\phi} \rangle = \sqrt{2 K_0} \begin{pmatrix}
        0_{4\times1}\\
        \alpha\\
        \tvec{\bar{k}}
    \end{pmatrix}
\end{equation}
with $\alpha = \sqrt{1 - \bar{\tvec{k}}^\trans \bar{\tvec{k}}}$ vanishing at a charge-conserving minimum.\\

Let us briefly recap how we get the mass spectrum in any THDM for the charge-conserving vacuum, that is, a vacuum respecting the observed electroweak symmetry breaking. First, we express the potential $V$ in terms of gauge-invariant bilinears employing~\eqref{eq:phiK}. From the potential
we compute ${\cal M} = \partial_\mu \partial_\nu V$. In turn we can compute $\widehat{\cal M}_{\text{neutral}}^2$ in~\eqref{eq:allOrderMneutral} with $\gamma_3$ given in~\eqref{eq:ga3def}. Diagonalising the matrix $\widehat{\cal M}_{\text{neutral}}^2$ according to~\eqref{eq:diagonalBasisNeutral} we get the scalar mass squared eigenvalues. The degenerate charged Higgs-boson masses squared follow directly from the minimum of the potential from~\eqref{eq:allOrderMcharged}.

We emphasise that we have not specified the explicit form of the potential~$V$ in this section. 
Indeed, it can be any THDM potential, 
for instance the tree-level potential~\eqref{eq:potK4} or a higher loop effective THDM potential. All the expressions given above in this section remain valid.
This holds in particular for the scalar mass spectrum
given in~\eqref{eq:allOrderMcharged} and~\eqref{eq:diagonalBasisNeutral}. The charged Higgs-boson mass squared is $m_{H^\pm}^2 = 4 u K_0$ for any THDM to any loop order -- a quite remarkable result: From the minimum of the potential, that is, from the solutions of~\eqref{eqbEW}-\eqref{eqEW}, we get $K_0$ and $u$ at the minimum directly giving the charged Higgs-boson mass squared.
These results illustrate the benefits of working with gauge invariants in general, and in particular in  studying THDMs using the bilinear formalism. We emphasise that we have derived the mass matrices in a completely gauge-invariant way.

\subsection{Tree-level scalar mass matrix}

Eventually, we would like to 
explicitly show our result 
in the case of a general THDM
tree-level potential. The charged scalar masses squared are
\begin{equation} \label{mHpm}
m_{H^\pm}^2 = 4 u K_0
\end{equation}
as discussed before. Furthermore, the neutral
$3 \times 3$ 
matrix~\eqref{eq:allOrderMneutral} becomes in this case
\begin{equation} \label{eq:MneutralTreeLevel}
    \widehat{\mathcal{M}}_\mathrm{neutral}^2 = 4 K_0 \left[ \eta_{00} \tvec{k} \tvec{k}^\trans + \tvec{\eta} \tvec{k}^\trans + \tvec{k} \tvec{\eta}^\trans + E + u (\unitmatrix_3 -  \tvec{k} \tvec{k}^\trans) \right].
\end{equation}
which is explicitly gauge invariant. 
This real symmetric matrix can be diagonalised
with the rotation matrix~$R$,  \eqref{eq:diagonalBasisNeutral}.
Under this change of basis, the bilinears transform as
shown in~\eqref{bilbasis} and the parameters of the tree-level potential transform as, \eqref{eq:b2},
\begin{equation} 
\label{eq:b2b}
\bar{\xi}_0 = \xi_0, \quad
\bar{\tvec{\xi}} = R \tvec{\xi}, \quad
\bar{\eta}_{00} = \eta_{00}, \quad
\bar{\tvec{\eta}} = R \tvec{\eta}, \quad
\bar{E} = R  E R^\trans \, .
\end{equation}
Therefore, the neutral mass matrix becomes
\begin{equation} \label{eq:Mbar0tree}
    \bar{\mathcal{M}}_\mathrm{neutral}^2 = 4 K_0 \left[ \eta_{00} \bar{\tvec{k}} \bar{\tvec{k}}^\trans + \bar{\tvec{\eta}} \bar{\tvec{k}}^\trans + \bar{\tvec{k}} \bar{\tvec{\eta}}^\trans + \bar{E} + u (\unitmatrix_3 -  \bar{\tvec{k}} \bar{\tvec{k}}^\trans) \right].
\end{equation}
Let us note that in practical calculations we can use~\eqref{eq:Mbar0tree} and \eqref{mHpm} together with the parameters $\eta_{00}$ and $\tvec{\bar{\eta}}$ to fix the parameter matrix~$\bar{E}$ of the THDM in terms of the scalar masses.
The scalar mass squared matrix for the tree level case, given in~\eqref{eq:Mbar0tree}, agrees with the known result in the Higgs basis (where $\bar{\tvec{k}}= \begin{pmatrix} 0, 0, 1 \end{pmatrix}^\trans)$; see for instance~\cite{Maniatis:2006fs}. In particular we get the squared masses of the scalar sector as known from the conventional formalism.


\section{Scalar trilinear and quartic interactions}
\label{sec:treeLeveLScalarCouplings}

Having found the scalar squared mass matrix in terms of gauge invariants in the last section, we can now proceed and express the scalar trilinear and quartic couplings in a gauge-invariant way. The connection between the bilinears and the component fields~\eqref{eq:doublets} is given in terms of the $\Gamma$ matrices~\eqref{eq:defgamma}.
From these $\Gamma$ matrices
we can express the cubic and quartic 
couplings in terms of the squared mass matrix~\eqref{eq:msExpr}. We emphasise that 
this matrix can be any THDM mass matrix, not restricted to the tree level THDM. 

We find for the cubic and quartic interactions in any THDM
\begin{align}
    \lambda_{ijk} &= \left(\partial_i M_s^2\right)^{jk} = \left(\Delta^\mu_{ij} \Gamma_k^\nu + \Delta^\mu_{ik} \Gamma_j^\nu + \Delta^\mu_{jk} \Gamma_i^\nu\right) \mathcal{M}_{\mu \nu}\,,\\
    \lambda_{ijkl} &= \left(\partial_i \partial_j M_s^2\right)^{kl} = \left(\Delta^\mu_{ij} \Delta^\nu_{kl} + \Delta^\mu_{ik} \Delta^\nu_{jl} + \Delta^\mu_{il} \Delta^\nu_{jk}\right) \mathcal{M}_{\mu \nu}\,.
\end{align}
In the mass basis~\eqref{eq:diagonalMs}, this yields
\begin{align}
    \bar{\lambda}_{ijk} &= \left(\bar{\Delta}^\mu_{ij} \bar{\Gamma}_k^\nu + \bar{\Delta}^\mu_{ik} 
    \bar{\Gamma}_j^\nu + \bar{\Delta}^\mu_{jk} \bar{\Gamma}_i^\nu\right) \bar{\mathcal{M}}_{\mu \nu}\,, \label{eq:massCubic}\\
    \bar{\lambda}_{ijkl} &= \left(\bar{\Delta}^\mu_{ij} \bar{\Delta}^\nu_{kl} + \bar{\Delta}^\mu_{ik} \bar{\Delta}^\nu_{jl} + \bar{\Delta}^\mu_{il} \bar{\Delta}^\nu_{jk}\right) \bar{\mathcal{M}}_{\mu \nu}\,. \label{eq:massQuartic}
\end{align}
The expressions of $\bar{\Delta}$ and $\bar{\Gamma}$ can be inferred from~\eqref{massbGD} in the last section. The matrix ${\mathcal M}$ is defined in~\eqref{eq:Mmunu} and
$\bar{\mathcal{M}}$ follows from ${\mathcal M}$ by~\eqref{eq:allOrderMneutral} and 
then~\eqref{eq:diagonalBasisNeutral}.

In order to get simple expressions, we first want to 
introduce new indices for the different scalars appearing in~\eqref{eq:diagonalMs}. 
For the $8\times8$ matrix~\eqref{eq:diagonalMs}, we have $\left(\bar{M}^2_s\right)_{ij}$ with $i,j \in \{1, \ldots, 8\}$. The entries $i,j = 1$ correspond to the neutral Goldstone mode $G^0$, the two following entries, that is, $i,j \in \{2, 3\}$ to the charged Goldstone modes $G^\pm$, whereas  $ i,j \in \{4, 5\}$ correspond to the charged Higgs bosons $H^\pm$, and the entries $i,j \in \{6,7,8\}$ to the three neutral scalars -- in the CP conserving case usually denoted by $h$, $H$, $A$ or $h^0$, $H^0$, $A^0$. 

We use in the following the index $G^0 = 1$ corresponding to $i=1$, the
indices $G^\pm_p, G^\pm_q \in \{1, 2\}$ referring to the second and third index $i,j,\ldots \in \{2, 3\}$. This means that the indices $G^\pm_p, G^\pm_q$ are shifted by one unit down with respect to the indices $i,j$.
Similar, for 
the charged Higgs bosons we use the indices $H^\pm_p, H^\pm_q \in \{1, 2\}$ referring to the original indices $\{4, 5\}$, that is, shifted by three units.
Eventually, we use the indices $s,t, \ldots \in \{1,2,3\}$ corresponding to $i,j,\ldots \in \{6,7,8\}$, hence, shifted five units.
 With this notation, we can write down all the non-vanishing components of the matrices $\bar{\Gamma}$ and $\bar{\Delta}$:
\begin{align}
    \bar{\Gamma}^0_s &= \sqrt{2 K_0} \, \bar{k}^s, & \bar{\Gamma}^a_s &= \sqrt{2 K_0} \, \delta^{as}, \label{eq:GammaDeltaExprs1}\\
    \bar{\Delta}^0_{st} &= \delta^{st}, & \bar{\Delta}^a_{st} &= \delta^{sa} \bar{k}^t + \delta^{ta} \bar{k}^s - \delta^{st} \bar{k}^a,\\
    \bar{\Delta}^\mu_{G^\pm_p G^\pm_q} &= \frac{\bar{K}^\mu}{K_0} \delta^{G^\pm_p G^\pm_q}, & \bar{\Delta}^\mu_{H^\pm_p H^\pm_q} &= \frac{\tilde{g}
    \overline{\KT}^\mu}{K_0} \delta^{H^\pm_p H^\pm_q}, \\
    \bar{\Delta}^\mu_{G^0 G^0} &= \frac{\bar{K}^\mu}{K_0}, & 
    \bar{\Delta}^a_{s G^0} &= - \varepsilon_{ast} \bar{k}^t\,.
    \label{eq:GDend}
\end{align}
We recall that the index $a \in\{1,2,3\}$ denotes the three spatial components of Minkowski-type vectors with indices $\mu, \nu, \ldots \in \{0,1,2,3\}$. Similar,
we have non-vanishing $\bar{\Delta}^\mu$ matrices mixing the charged Higgs and Goldstone components. The corresponding entries are not uniquely fixed due to the possibility of rotating the corresponding real components (\textit{i.e.}~performing phase redefinitions of the charged fields). However, one can always write
\begin{equation} \label{eq:GammaDeltaExprs2}
    \bar{\Delta}^a_{G^\pm_p H^\pm_q} = \begin{pmatrix}
        x_a & -y_a\\
        y_a & x_a
    \end{pmatrix}^{G^\pm_p H^\pm_q} \equiv \left(\chi_a\right)^{G^\pm_p H^\pm_q}\,,
\end{equation}
where the 3-vectors $\tvec{x}$ and $\tvec{y}$ are defined such that $(\tvec{x}, \tvec{y}, \tvec{\bar{k}})$ constitutes an orthonormal basis of the 3-dimensional bilinear field space, \textit{i.e.}
\begin{equation} \label{eq:xykRelations}
    \tvec{x}^\trans \tvec{\bar{k}} = \tvec{y}^\trans \tvec{\bar{k}} = \tvec{x}^\trans \tvec{y} = 0 \quad \text{and} \quad \tvec{x}^\trans \tvec{x} = \tvec{y}^\trans \tvec{y} = \tvec{\bar{k}}^\trans \tvec{\bar{k}} = 1\,,
\end{equation}
oriented such that $\tvec{x}\times\tvec{y} = \tvec{\bar{k}}$. In fact, $\tvec{x}$ and $\tvec{y}$ characterise the orthogonal transformation $R_H$ allowing to rotate $\tvec{\bar{k}}$ to the Higgs basis which was used in App.~\ref{app:scalarMassMatrixStructure} to diagonalise the mass matrix. Namely, one has
\begin{equation}
    R_H = \begin{pmatrix}
        - \tvec{y}^\trans\\
        \tvec{x}^\trans\\
        \tvec{\bar{k}}^\trans
    \end{pmatrix} = 
    \begin{pmatrix}
        -y_1 & -y_2 & -y_3\\
        x_1 & x_2 & x_3\\
        \bar{k}_1 & \bar{k}_2 & \bar{k}_3
    \end{pmatrix}, \qquad R_H \tvec{k} = \begin{pmatrix}
        0 \\ 0 \\ 1
    \end{pmatrix}\,,
\end{equation}
which, from~\eqref{eq:xykRelations} is manifestly orthogonal. In turn, the reason why $\tvec{x}$ and $\tvec{y}$ are not uniquely defined is clear: there exist infinitely many orthogonal transformations $R_H'$ rotating $\tvec{k}$ to the Higgs basis, related by arbitrary rotations around the $z$-axis. 

While the matrices $\chi_a$ are themselves not uniquely defined, it is useful to observe that since
\begin{equation}
    \tvec{x} \tvec{x}^\trans + \tvec{y} \tvec{y}^\trans + \tvec{\bar{k}} \tvec{\bar{k}}^\trans = \unitmatrix_3 
    \quad \text{and} \quad \tvec{x} \times \tvec{y} = \tvec{\bar{k}}\,,
\end{equation}
we obtain
\begin{equation}
    x_a x_b + y_a y_b = \delta_{ab} - \bar{k}_a \bar{k}_b \quad \text{and} \quad x_a y_b - x_b y_a = \varepsilon_{abc} \bar{k}^c\,.
\end{equation}
Therefore, for any two matrices $\chi_a$ and $\chi_b$, one has
\begin{equation}
    \chi_a \chi_b^\trans = \begin{pmatrix}
        \delta_{ab} - \bar{k}_a \bar{k}_b        & \varepsilon_{abc} \bar{k}^c\\\
        -\varepsilon_{abc} \bar{k}^c & \delta_{ab} - \bar{k}_a \bar{k}_b
    \end{pmatrix}\,,
\end{equation}
independently of the basis chosen to express $\tvec{x}$ and $\tvec{y}$.

\subsection{Tree level scalar cubic and quartic interactions}

Eventually, we would like to give the explicit expressions for the tree-level cubic and quartic interactions
in gauge-invariant form. These interactions are derived from the tree-level potential.

The neutral scalar mass matrix in the mass basis has been computed already in~\eqref{eq:Mbar0tree}. Together with
the expressions~\eqref{eq:GammaDeltaExprs1}--\eqref{eq:GammaDeltaExprs2}
we get the cubic and quartic interactions~\eqref{eq:massCubic}, and~\eqref{eq:massQuartic}, respectively.
It turns out to be convenient to define the 
following 3-vectors:
\begin{align}
    \bar{f}^a &= 8 K_0 \left(\eta_{00} \bar{k}^a + \bar{\eta}^a\right) - \bar{k}^a m_a^2\,,\\
    \bar{f}_\pm^a &= 8 K_0 \left[\left(\eta_{00} - u\right) \bar{k}^a + \bar{\eta}^a\right] = 8 K_0 \left(\eta_{00} \bar{k}^a + \bar{\eta}^a\right) - 2 \bar{k}^a m_{H^\pm}^2 \,.
\end{align}
Let us note that an index appearing with a mass does not imply summation, for instance in the expression $\bar{k}^a m_a^2$.

The part of the Lagrangian containing the scalar cubic interactions can be written, after spontaneous symmetry breaking and rotation to the mass basis, as
\begin{align}
\begin{split}
    - \mathcal{L} \supset & \frac{1}{3!} \bar{\lambda}_{stu} \, h^s h^t h^u + \frac{1}{2} \bar{\lambda}_{stG^0} \, h^s h^t G^0 + \frac{1}{2} \bar{\lambda}_{sG^0G^0} \, h^s G^0 G^0\\
    &+ \bar{\lambda}_{s{H^\pm}{H^\pm}} \, h^s H^- H^+ + \bar{\lambda}_{s{G^\pm}{G^\pm}} \, h^s G^- G^+ + \left[\bar{\lambda}_{s{G^\pm}{H^\pm}} \, h^s G^- H^+ + \mathrm{h.c.} \right]\,.
\end{split}
\end{align}
The complex charged fields $H^\pm$, $G^\pm$ are given by
\begin{align}
    H^\pm &= \frac{e^{\pm i \omega_H}}{\sqrt{2}} \left(H^\pm_1 \pm i H^\pm_2 \right)\,,
    \\
    G^\pm &= \frac{e^{\pm i \omega_G}}{\sqrt{2}} \left(G^\pm_1 \pm i G^\pm_2 \right)\,,
\end{align}
where the inclusion of $\omega_H$ and $\omega_G$ reflects the possibility of performing arbitrary phase redefinitions of the complex fields. The analytic expressions for the various cubic couplings are readily derived through~\eqref{eq:massCubic}, \eqref{eq:GammaDeltaExprs1}--\eqref{eq:GammaDeltaExprs2} and put into the form
\begin{equation}
\begin{split}
    &\bar{\lambda}_{stu} = \frac{1}{\sqrt{2 K_0}}\bigg\{\left(\delta^{st} - \bar{k}^s \bar{k}^t\right) \bar{f}_\pm^u + \delta^{st} \bar{k}^u \left( m_s^2 + m_t^2 - m_u^2\right)  \bigg\} + (s \leftrightarrow u) + (t \leftrightarrow u)\,,
    \\
    &\bar{\lambda}_{st G^0} = \varepsilon_{stu} \frac{m_t^2 - m_s^2}{\sqrt{2 K_0}}  \bar{k}^u\,,
    \\
    &\bar{\lambda}_{s G^0 G^0} = \frac{m_s^2}{\sqrt{2 K_0}} \bar{k}^s\,,
    \\
    &\bar{\lambda}_{s H^\pm H^\pm} = \frac{\bar{f}^s}{\sqrt{2 K_0}} \label{eq:cubicNeutralCharged}\,,
    \\
    &\bar{\lambda}_{s G^\pm G^\pm} = \frac{m_s^2}{\sqrt{2 K_0}} \bar{k}^s\,,
    \\
    &\bar{\lambda}_{s G^\pm H^\pm} = \frac{m_s^2 - m_{H^\pm}^2}{\sqrt{2 K_0}} e^{i(\omega_G - \omega_H)}\left(x^s + i y^s\right)\,.
\end{split}
\end{equation}
From the last expression, we see that a phase redefinition involving $\omega_G - \omega_H$ can always be compensated by a redefinition of the vectors $\tvec{x}$ and $\tvec{y}$ already mentioned above. More precisely, by defining a new pair of vectors $\tvec{x}'$ and $\tvec{y}'$ such that
\begin{equation} \label{eq:xyPhaseRedef}
    \begin{pmatrix}
        x'_i \\
        y'_i
    \end{pmatrix} = \begin{pmatrix}
        \cos\left(\omega_G - \omega_H\right) & -\sin\left(\omega_G - \omega_H\right)\\
        \sin\left(\omega_G - \omega_H\right) & \cos\left(\omega_G - \omega_H\right)
    \end{pmatrix} \begin{pmatrix}
        x_i \\
        y_i
    \end{pmatrix}\,,
\end{equation}
the $h_s G^\pm H^\pm$ vertex is simply rewritten as
\begin{equation}
    \bar{\lambda}_{s G^\pm H^\pm} = \frac{m_s^2 - m_{H^\pm}^2}{\sqrt{2 K_0}}\left(x'_s + i y'_s\right)\,.
\end{equation}
Therefore, here and in the following, we shall for clarity drop the arbitrary phases $\omega_{G,H}$ from the expressions of the tree-level vertices without any loss of generality. Instead, one should keep in mind that vertices involving two distinct charged fields can only be defined up to unphysical phase redefinitions.\\

The part of the Lagrangian corresponding to the scalar quartic interactions reads
{\allowdisplaybreaks
\begin{align}
\begin{split}
    - \mathcal{L} &\supset \frac{1}{4!} \bar{\lambda}_{stuv} \, h^s h^t h^u h^v+ \frac{1}{3!} \bar{\lambda}_{stuG^0} \, h^s h^t h^u G^0 + \frac{1}{4} \bar{\lambda}_{stG^0G^0} \, h^s h^t G^0 G^0\\
    &+ \frac{1}{3!} \bar{\lambda}_{sG^0G^0G^0} \, h^s G^0 G^0 G^0  + \frac{1}{4!} \bar{\lambda}_{G^0G^0G^0G^0} \, G^0 G^0 G^0 G^0\\
    & + \frac{1}{2} \bar{\lambda}_{st H^\pm H^\pm} \, h^s h^t H^- H^+  + \bar{\lambda}_{s G^0 H^\pm H^\pm} \, h^s G^0 H^- H^+  + \frac{1}{2} \bar{\lambda}_{G^0 G^0 H^\pm H^\pm} \, G^0 G^0 H^- H^+\\
    & + \frac{1}{2} \bar{\lambda}_{st G^\pm G^\pm} \, h^s h^t G^- G^+  + \bar{\lambda}_{s G^0 G^\pm G^\pm} \, h^s G^0 G^- G^+  + \frac{1}{2} \bar{\lambda}_{G^0 G^0 G^\pm G^\pm} \, G^0 G^0 G^- G^+\\
    & + \left[ \frac{1}{2} \bar{\lambda}_{st G^\pm H^\pm} \, h^s h^t G^- H^+  + \bar{\lambda}_{s G^0 G^\pm H^\pm} \, h^s G^0 G^- H^+  + \frac{1}{2} \bar{\lambda}_{G^0 G^0 G^\pm H^\pm} \, G^0 G^0 G^- H^+ + \mathrm{h.c.}\right]\\
    &+ \frac{1}{4} \bar{\lambda}_{H^\pm H^\pm H^\pm H^\pm} \, H^- H^+ H^- H^+ + \bar{\lambda}_{G^\pm G^\pm H^\pm H^\pm} \, G^- G^+ H^- H^+ \\
    &+ \frac{1}{4} \bar{\lambda}_{G^\pm G^\pm G^\pm G^\pm} \, G^- G^+ G^- G^+\\
    &+ \left[\frac{1}{2}\bar{\lambda}_{G^\pm H^\pm H^\pm H^\pm} \, G^- H^+ H^- H^+ + \frac{1}{2}\bar{\lambda}_{G^\pm G^\pm G^\pm H^\pm} \, G^- G^+ G^- H^+ + \mathrm{h.c.} \right]\,.
\end{split}
\end{align}
}
The analytic expressions for the various quartic couplings in gauge-invariant form can be derived 
using Eqs.~\eqref{eq:massQuartic}, \eqref{eq:GammaDeltaExprs1}--\eqref{eq:GammaDeltaExprs2} and are given in App.~\ref{sub:scalarint}.


\section{Gauge sector}
\label{sec:oneLoopGauge}

\subsection{Definitions}

Turning to the gauge sector, we are interested in expressing all relevant quantities (gauge boson masses and vertices) in terms of gauge-invariant quantities. The gauge generators of the $8$-component scalar multiplet $\phi$ under $\eweakgroup$ are noted $T^\phi_A$, $A = 1,\dots,4$, where $T_1^\phi$ stands for the $U(1)_Y$ generator while $T_{2,3,4}^\phi$ correspond to $\mathit{SU}(2)_L$ transformations. The kinetic part of the Lagrangian which couples  the scalar fields to the gauge bosons is given, before symmetry breaking, by\footnote{Here and in the following, we use a metric of signature $(+,-,-,-)$.}
\begin{equation} \label{eq:scalarGaugeLag}
    \mathcal{L}\supset \frac{1}{2}  D^\mu \phi_i D_\mu \phi^i
\end{equation}
where the scalar field covariant derivative can be generically written as
\begin{equation}
    D_\mu \phi^i = \partial_\mu \phi^i + i V_\mu^A G^{AB} \left(T^\phi_B\right)^{ij} \phi^j\,.
\end{equation}
In the above expression $V_\mu^A$ stands for the vector fields whereas $G$ is a $4\times4$ diagonal matrix populated with the gauge couplings $g_1$ and $g_2$ according to
\begin{equation}
    G = \diag\left(g_1, g_2, g_2, g_2\right)\,.
\end{equation}
Expanding \eqref{eq:scalarGaugeLag} yields
\begin{align}
    \mathcal{L} &\supset \frac{1}{2} \partial^\mu \phi_i \partial_\mu \phi^i - i V^\mu_A G^{AB} \left(T^\phi_B\right)^{ij} \phi^i \partial_\mu \phi^j + \frac{1}{4} V_\mu^A V^\mu_B G^{AC} G^{BD} \left\{T^\phi_C, T^\phi_D\right\}^{ij} \phi^i \phi^j\,.
\end{align}
After spontaneous symmetry breaking, the scalar multiplet is shifted according to \begin{equation}
    \phi \rightarrow \langle \phi \rangle + \omega
\end{equation}
such that the Lagrangian contains in particular the following interactions
\begin{equation} \label{eq:LbrokenGauge}
    \mathcal{L}_\mathrm{broken} \supset \frac{1}{2} \left(M_G^2\right)^{AB} V_\mu^A V^\mu_B + g_{Aij} V^\mu_A \omega^i \partial_\mu \omega^j + \frac{1}{2} g_{ABi} V_\mu^A V^\mu_B \omega^i + \frac{1}{4} g_{ABij} V_\mu^A V^\mu_B \omega^i \omega^j\,,
\end{equation}
where the gauge boson mass matrix and the vector-scalar-scalar (VSS), vector-vector-scalar (VVS) and vector-vector-scalar-scalar (VVSS) couplings are respectively given by\footnote{We have defined the couplings so that they match the conventions in 
\cite{Martin:2017lqn}, despite a different convention for the Minkowski metric.}
\begin{align}
    \left(M_G^2\right)^{AB} &= \frac{1}{2} G^{AC} G^{BD} \left\{T^\phi_C, T^\phi_D\right\}^{ij} \langle\phi\rangle^i \langle\phi\rangle^j\,,\\
    g_{Aij} &= - i G^{AB} \left(T^\phi_B\right)^{ij}\,, \label{eq:gVSS}\\
    g_{ABi} &= G^{AC} G^{BD} \left\{T^\phi_C, T^\phi_D\right\}^{ij} \langle\phi\rangle^j \,, \label{eq:gVVS}\\
    g_{ABij} &= G^{AC} G^{BD} \left\{T^\phi_C, T^\phi_D\right\}^{ij}\, . \label{eq:gVVSS}
\end{align}

\subsection{Gauge-boson mass matrix}

With the definitions of the previous subsection let us first examine the gauge-boson mass matrix $M_G^2$. With standard conventions for the gauge symmetry generators, the entries of $M_G^2$ cannot generally be expressed in terms of bilinear fields in a basis-independent way. On the other hand, its matrix invariants -- and in particular its eigenvalues -- can be expressed in such a way. From a direct calculation, it is straightforward to show that $M_G^2$ can be rotated into the diagonal form
\begin{equation}
    \bar{M}_G^2 = \diag\left(M_\gamma^2, M_W^2, M_W^2, M_Z^2\right)\,,
\end{equation}
where the eigenvalues are given in terms of the bilinear fields by
\begin{equation}\label{eq:gaugeEigs}
	M_W^2 = \frac{1}{2}  \left(g_+^2 - g_-^2\right) K_0= \frac{1}{2} g_2^2 K_0, \quad 
	M^{2}_{Z, \gamma} = \frac{1}{2}\left[g_{+}^2 K_0 \pm \sqrt{g_{+}^4 \tvec{K}^2  + g_{-}^4 (K_0^2-\tvec{K}^2)}\,\right]\,,
\end{equation}
with
\begin{equation}
	g_+^2 \equiv \frac{g_1^2 + g_2^2}{2} \quad \text{and} \quad g_-^2 \equiv \frac{g_1^2 - g_2^2}{2}\,.
\end{equation}
At a charge-conserving minimum where $K_0^2 = \tvec{K}^2$, we obtain from \eqref{eq:gaugeEigs} the familiar result
\begin{equation}
    M_Z^2 = g_+^2 K_0 = \frac{1}{2}\left(g_1^2 + g_2^2\right) K_0, \qquad
    M_\gamma = 0\,.
\end{equation}

In what follows, we are only interested in the physically relevant case of a charge-conserving minimum where $K_0^2 = \tvec{K^2}$, that is a spontaneous symmetry breaking of the type $\eweakgroup \rightarrow \emgroup$. In that case, the gauge-boson mass matrix in the original basis (\textit{i.e.}~before diagonalisation) conveniently reduces to
\begin{equation}
    M_G^2 = \frac{K_0}{2} \begin{pmatrix}
        g_1^2 & 0 & 0 & -g_1 g_2 \\
        0 & g_2^2 & 0 & 0 \\
        0 & 0 & g_2^2 & 0 \\
        -g_1 g_2 & 0 & 0 & g_2^2
    \end{pmatrix}
\end{equation}
and is diagonalised according to
\begin{equation}
    \bar{M}_G^2 = R_W M_G^2 R_W^\trans = \diag\left(0, M_W^2, M_W^2, M_Z^2\right)\,,
\end{equation}
where
\begin{equation} \label{eq:gaugeRW}
    R_W = \begin{pmatrix}
    \cos\theta_W & 0 & 0 & \sin\theta_W \\
    0 & \ 1 \  & 0 & 0 \\
    0 & 0 & \ 1 \ & 0 \\
    -\sin\theta_W & 0 & 0 & \cos\theta_W
    \end{pmatrix}\,,
\end{equation}
with $\theta_W$ the weak-mixing angle, satisfying
\begin{equation}
    \cos \theta_W = \frac{g_2}{\sqrt{g_1^2 + g_2^2}} = \frac{M_W}{M_Z}\,.
\end{equation}

\subsection{Scalar generators in the mass basis}
\label{sec:scalar-gen}

While expressing the gauge-boson mass matrix in terms of bilinear fields at a charge-conserving minimum was a trivial task, the derivation of the scalar-vector vertices/couplings in the physical basis is considerably more involved. With view on~\eqref{eq:gVSS}--\eqref{eq:gVVSS}, we see that such a computation reduces in fact to the determination of the expression of the gauge generators in the mass basis. Recalling that the bilinear fields are given by
\begin{equation}
    K^\mu = \frac{1}{2} \Delta^\mu_{ij} \phi^i \phi^j\,,
\end{equation}
the invariance of $K^\mu$ under infinitesimal gauge transformations of the form
\begin{equation}
    \phi^i \rightarrow \left(\unitmatrix_8 + i \theta^A T^\phi_A\right)^{ij} \phi^j
\end{equation}
translates, for all $\mu \in \{0,\dots,3\}$ and $A\in \{1,\dots,4\}$, into the condition
\begin{equation} \label{eq:deltaTcommOriginalBasis}
    \left[\Delta^\mu, T^\phi_A\right]^{ij} = 0\,.
\end{equation}
Clearly, the above relation remains valid after rotation of the scalar multiplet. In particular, defining the scalar generators $\bar{T}^\phi$ rotated to the mass basis, \eqref{eq:deltaTcommOriginalBasis} translates to
\begin{equation} \label{eq:deltaTcommMassBasis}
    \left[\bar{\Delta}^\mu, \bar{T}^\phi_A\right]^{ij} = 0\,.
\end{equation}
With the analytic expressions of $\bar{\Delta}^\mu$ in terms of bilinear fields at hand, the derivation of $\bar{T}^\phi_A$ simply amounts to solving a system of linear equations. Doing so yields a four-dimensional basis of purely imaginary skew-symmetric matrices, noted $\left\{\Theta_B\right\}$, such that
\begin{equation}
    \left(\bar{T}^\phi_A\right)^{ij} = X^{AB} \Theta_B^{ij}\,.
\end{equation}
The matrix of coefficients $X^{AB}$ is most easily fixed by considering a set of quantities independent of the basis chosen for the scalar multiplet. For instance, requiring that the quantity
\begin{equation}
    \Gamma^\mu_i \Gamma^\nu_j \left(T^\phi_A\right)^{ij} = \Gamma^\mu\, T^\phi_A\, {\Gamma^\nu}^T
\end{equation}
remains unchanged (for all indices $\mu$, $\nu$, $A$) after rotation to the mass basis provides a set of linear constraints on the coefficients
$X^{AB}$. Requiring in addition the gauge boson mass matrix $M_G^2$ to be invariant under a change of basis allows to uniquely determine the expressions of $\bar{T}^\phi_A$. Before presenting the final expression, it is useful to define two generators $\bar{T}^\phi_\pm$ such that
\begin{equation} \label{eq:T23vsTpm}
    \begin{pmatrix}
        \bar{T}^\phi_2 \\
        \bar{T}^\phi_3
    \end{pmatrix} = \begin{pmatrix}
        \cos\theta & -\sin\theta \\
        \sin\theta & \cos\theta
    \end{pmatrix} \begin{pmatrix}
        \bar{T}^\phi_+ \\
        \bar{T}^\phi_-
    \end{pmatrix} = \begin{pmatrix}
        \cos\theta \, \bar{T}^\phi_+ -\sin\theta \, \bar{T}^\phi_- \\
        \sin\theta \, \bar{T}^\phi_+ + \cos\theta \, \bar{T}^\phi_-
    \end{pmatrix}\,,
\end{equation}
where the precise value of $\theta$ depends on the basis chosen to express the vectors $\tvec{x}$ and $\tvec{y}$ (we will come back to this point further below). We may now provide the main result of this section, namely the expression of the gauge generators $\bar{T}^\phi_{1,4}$ and $\bar{T}^\phi_\pm$ in the mass basis. It is useful to reiterate that, in this basis, the dynamical field multiplet reads
\begin{equation}
    \bar{\varphi} = \begin{pmatrix} G^0, & G^\pm_1, & G^\pm_2, & H^\pm_1, & H^\pm_2, & h_1^0, & h_2^0, & h_3^0 \end{pmatrix}\,.
\end{equation}
We have:
\begin{align}
    \bar{T}^\phi_1 &= \frac{1}{2}\begin{pmatrix}
        0 & \begin{matrix} 0 
        & 0\end{matrix} 
        & \begin{matrix} 0 & 0\end{matrix}
        & i\tvec{\bar{k}}^\trans \\
        \begin{matrix} 0\\ 0 \end{matrix} & \sigma_2 & 
        \begin{matrix} 0 & 0\\ 0 & 0\end{matrix} & 
        0_{2\times 3} \\
        \begin{matrix} 0\\ 0 \end{matrix} & \begin{matrix} 0 & 0\\ 0 & 0\end{matrix} 
        & \sigma_2 & 0_{2\times 3} \\
        -i \tvec{\bar{k}} 
        & 0_{3\times 2} 
        & 0_{3\times 2} & i \Lambda_{\tvec{\bar{k}}}
    \end{pmatrix}, &
    \bar{T}^\phi_+ &= \frac{1}{2}\begin{pmatrix}
        0 
        & \begin{matrix} 0 & i\end{matrix} 
        & \begin{matrix}0 & 0 \end{matrix} & 0_{1\times 3} \\
        \begin{matrix} 0\\ -i\end{matrix}
        &\begin{matrix} 0 & 0\\ 0 &     0\end{matrix}
        &\begin{matrix} 0 & 0\\ 0 & 0\end{matrix}
        &\begin{matrix} -i \tvec{\bar{k}}^\trans\\ 
            0_{1\times 3}\end{matrix}\\
        \begin{matrix} 0\\0\end{matrix}
        &\begin{matrix} 0 & 0\\ 0 & 0\end{matrix}
        &\begin{matrix} 0 & 0\\ 0 & 0\end{matrix}
        &\begin{matrix} i\tvec{y}^\trans\\
            i \tvec{x}^\trans\end{matrix}\\
        0_{3\times 1} 
        & \begin{matrix} i \tvec{\bar{k}} & 0_{3\times 1}\end{matrix}
        &\begin{matrix} -i \tvec{y} & -i \tvec{x} \end{matrix}
        &0_{3\times 3}
    \end{pmatrix}, \\
    \bar{T}^\phi_4 &= \frac{1}{2}\begin{pmatrix}
        0 & \begin{matrix} 0 
        & 0\end{matrix} 
        & \begin{matrix} 0 & 0\end{matrix}
        & -i\tvec{\bar{k}}^\trans \\
        \begin{matrix} 0\\ 0 \end{matrix} & \sigma_2 & 
        \begin{matrix} 0 & 0\\ 0 & 0\end{matrix} & 
        0_{2\times 3} \\
        \begin{matrix} 0\\ 0 \end{matrix} & \begin{matrix} 0 & 0\\ 0 & 0\end{matrix} 
        & \sigma_2 & 0_{2\times 3} \\
        i \tvec{\bar{k}} 
        & 0_{3\times 2} 
        & 0_{3\times 2} & -i \Lambda_{\tvec{\bar{k}}}
    \end{pmatrix}, &
    \bar{T}^\phi_- &= \frac{1}{2}\begin{pmatrix}
        0 
        & \begin{matrix} -i & 0\end{matrix} 
        & \begin{matrix}0 & 0 \end{matrix} & 0_{1\times 3} \\
        \begin{matrix} i\\ 0\end{matrix}
        &\begin{matrix} 0 & 0\\ 0 &     0\end{matrix}
        &\begin{matrix} 0 & 0\\ 0 & 0\end{matrix}
        &\begin{matrix} 0_{1\times 3}\\ -i \tvec{\bar{k}}^\trans 
            \end{matrix}\\
        \begin{matrix} 0\\0\end{matrix}
        &\begin{matrix} 0 & 0\\ 0 & 0\end{matrix}
        &\begin{matrix} 0 & 0\\ 0 & 0\end{matrix}
        &\begin{matrix} -i\tvec{x}^\trans\\
            i\tvec{y}^\trans\end{matrix}\\
        0_{3\times 1} 
        & \begin{matrix} 0_{3\times 1} & i \tvec{\bar{k}} \end{matrix}
        &\begin{matrix} i \tvec{x} & -i \tvec{y} \end{matrix}
        &0_{3\times 3}
    \end{pmatrix}\,,
\end{align}
where $0_{m\times n}$ is the $m \times n$ zero matrix
and where the auxiliary matrix $\Lambda_{\tvec{k}}$ is given by
\begin{equation}
    \left(\Lambda_{\tvec{\bar{k}}}\right)^{ij} = \varepsilon_{ijk} \bar{k}^k\,.
\end{equation}
A few remarks are in order. First, we may check that the generator of the residual gauge group $\emgroup$ is given by
\begin{equation}
        \bar{T}^\phi_\mathrm{em} = \bar{T}^\phi_1 + \bar{T}^\phi_{4} = \begin{pmatrix}
        0 &  &  &  \\
         & \sigma_2 &  & \\
         &  & \sigma_2 &  \\
         &  &  & 0_{3\times 3}
    \end{pmatrix}\,,
\end{equation}
as expected from the fact that only $G^\pm$ and $H^\pm$ are electrically charged. Then, the presence of $\tvec{x}$ and $\tvec{y}$ in the expressions of $\bar{T}^\phi_\pm$ (and, hence, of $\bar{T}^\phi_{2,3}$) indicates an unphysical dependence of the gauge generators on the basis chosen to express these vectors (see also the discussion in Sec.~\ref{sec:treeLeveLScalarCouplings}). A similar statement holds regarding the angle $\theta$ introduced in~\eqref{eq:T23vsTpm}. More precisely, if we apply a phase shift to every charged field in the scalar-gauge sector,
\begin{align}
\begin{split}
    W^\pm &\rightarrow e^{i \omega_W} W^\pm\,,\\
    H^\pm &\rightarrow e^{i \omega_H} H^\pm\,,\\
    G^\pm &\rightarrow e^{i \omega_G} G^\pm\,,
\end{split}
\end{align}
the analytic expression of $\bar{T}^\phi_{2,3}$ remains unchanged if $\tvec{x}$, $\tvec{y}$ are redefined according to~\eqref{eq:xyPhaseRedef} and $\theta$ according to
\begin{equation}
    \theta \rightarrow \theta' = \theta - \omega_G - \omega_W\,.
\end{equation}
In other words, $\theta$ is arbitrary and may always be absorbed by a phase redefinition of $G^\pm$ and/or $W^\pm$. Hence, we shall take $\theta=0$ in the forthcoming analytic expressions without any loss of generality.\\

As a final remark, we note that the generators $(\bar{T}^\phi_A)^{ij}$ presented above have only been rotated to the basis of the scalar mass eigenstates. To obtain in addition their expression in the basis of the gauge mass eigenstates, one only needs to apply the rotation $R_W$, defined in~\eqref{eq:gaugeRW}, to the gauge index:
\begin{equation}
    (\bar{T}^\phi_A)^{ij} \rightarrow R_W^{AB} (\bar{T}^\phi_B)^{ij}\,.
\end{equation}

\subsection{Vector-scalar interactions}

Having determined the expression of the scalar generators in the mass basis, it is straightforward to compute the expression of all vector-scalar interaction vertices at tree-level in terms of the bilinear fields, 
using~\eqref{eq:gVSS}--\eqref{eq:gVVSS}. The part of the Lagrangian density describing the vector-scalar interactions reads
\begin{equation}
    \mathcal{L}_{V-S} = \mathcal{L}_{VSS} + \mathcal{L}_{VVS} + \mathcal{L}_{VVSS}\,,
\end{equation}
where, from~\eqref{eq:LbrokenGauge},
\begin{align}
    \mathcal{L}_{VSS} &= g_{Aij} V^\mu_A \varphi^i \partial_\mu \varphi^j\,,\\
    \mathcal{L}_{VVS} &= \frac{1}{2} g_{ABi} V_\mu^A V^\mu_B \varphi^i\,,\\
    \mathcal{L}_{VVSS} &= \frac{1}{4} g_{ABij} V_\mu^A V^\mu_B \varphi^i \varphi^j\,.
\end{align}
Applying~\eqref{eq:gVSS}--\eqref{eq:gVVSS} (the expectation value of the scalar multiplet in the mass basis can be read from~\eqref{eq:phiVacuum}) and retaining only non-zero couplings gives
\begin{align} \nonumber
    \mathcal{L}_{VSS} &= 
g_{s G^0 Z} h^s \partial_\mu G^0 Z^\mu
+
g_{H^\pm H^\pm Z} H^-  \partial_\mu H^+ Z^\mu
+
g_{G^\pm G^\pm Z} G^-  \partial_\mu G^+ Z^\mu
\\ & \nonumber
+
g_{H^\pm H^\pm \gamma} H^-  \partial_\mu H^+ A^\mu
+
g_{G^\pm G^\pm \gamma} G^-  \partial_\mu G^+ A^\mu
\\ &
+
g_{s H^\pm W^\pm} h^s \partial_\mu H^+ W^{- \mu}
+
g_{s G^\pm W^\pm} h^s \partial_\mu G^+ W^{- \mu}
+
g_{G^0 G^\pm W^\pm} G^0 \partial_\mu G^+ W^{- \mu}\,,
\end{align}

\begin{align}
    \mathcal{L}_{VVS} &= 
\frac{1}{2} g_{s Z Z} h^s Z^\mu Z_\mu
+
g_{s W^\pm W^\pm} h^s W^{- \mu} W^+_\mu
+
g_{G^\pm Z W^\pm} G^- Z^\mu W^+_\mu
+
g_{G^\pm \gamma W^\pm} G^- A^\mu W^+_\mu\,,
\end{align}

\begin{align}
    \mathcal{L}_{VVSS} &= \nonumber
\frac{1}{4} g_{s t Z Z} h^s h^t Z^\mu Z_\mu
+
\frac{1}{2} g_{s t W^\pm W^\pm} h^s h^t W^{- \mu} W^+_\mu
+
g_{s H^\pm Z W^\pm} h^s H^- Z^\mu W^+_\mu
\\ & \nonumber
+
g_{s G^\pm Z W^\pm} h^s G^- Z^\mu W^+_\mu
+
g_{s H^\pm \gamma W^\pm} h^s H^- A^\mu W^+_\mu
+
g_{s G^\pm \gamma W^\pm} h^s G^- A^\mu W^+_\mu
\\ & \nonumber
+
\frac{1}{4} g_{G^0 G^0 Z Z} G^0 G^0 Z^\mu Z_\mu
+
\frac{1}{2} g_{G^0 G^0 W^\pm W^\pm} G^0 G^0 W^{- \mu} W^+_\mu
+
g_{G^0 G^\pm Z W^\pm} G^0 G^- Z^\mu W^+_\mu
\\ & \nonumber
+
g_{G^0 G^\pm \gamma W^\pm} G^0 G^- A^\mu W^+_\mu
+
\frac{1}{2} g_{H^\pm H^\pm Z Z} H^- H^+ Z^\mu Z_\mu
+
\frac{1}{2} g_{G^\pm G^\pm Z Z} G^- G^+ Z^\mu Z_\mu
\\ & \nonumber
+
g_{H^\pm H^\pm Z \gamma} H^- H^+ Z^\mu A_\mu
+
g_{G^\pm G^\pm Z \gamma} G^- G^+ Z^\mu A_\mu
+
g_{H^\pm H^\pm \gamma \gamma} H^- H^+ A^\mu A_\mu
\\ & \nonumber
+
g_{G^\pm G^\pm \gamma \gamma} G^- G^+ A^\mu A_\mu
+
g_{H^\pm H^\pm W^\pm W^\pm} H^- H^+ W^{- \mu} W^+_\mu
\\ &
+
g_{G^\pm G^\pm W^\pm W^\pm} G^- G^+ W^{- \mu} W^+_\mu\,.
\end{align}
All couplings can be found in App.~\ref{sub:vectorscalar}.


\section{Yukawa sector}
\label{sec:yukawa}

\subsection{A basis-invariant description}

While the bilinear formalism offers, for any type of THDM, a compact and elegant basis-independent formulation of the scalar interactions, it is a priori unable to describe the scalar-fermion interactions which are by construction linear in the scalar fields. The purpose of this section is to show how such a basis-independent formalism can in fact be extended to the Yukawa sector.
Combined with the gauge-invariant approach developed in this work, we will be able to derive every Yukawa coupling in terms of gauge-invariant quantities, thus providing a complete description of the interactions among all physical states (scalars, vectors and fermions) in the most general THDM.\\

The part of the Lagrangian describing the Yukawa interactions involving the two Higgs doublets $\varphi_1$ and $\varphi_2$ can be first parameterised in the standard way:
\begin{align}
    -\mathcal{L}_Y = \Big[
    \overline{Q}_L\big( y_u\,\widetilde{\varphi}_1 + \epsilon_u\, \widetilde{\varphi}_2 \big) u_R
    + \overline{Q}_L\big( y_d\,\varphi_1 + \epsilon_d\, \varphi_2 \big) d_R
    + \overline{L}\big( y_e\,\varphi_1 + \epsilon_e\, \varphi_2 \big) e_R  
\Big] + \mathrm{h.c.}\,. \label{eq:yukLag}
\end{align}
As usual, $Q_L$ denotes the left-handed quark doublets and $L$ the left handed lepton doublets,
$u_R$, $d_R$ are the right-handed up- and down-type quark singlets, and $e_R$ the 
right-handed leptons. The corresponding Yukawa coupling matrices are denoted by $y_u$, $y_d$, $y_e$, as well as
$\epsilon_u$, $\epsilon_d$, $\epsilon_e$.
The conjugate doublets $\widetilde{\varphi}_a$ are given, as usual, by
\begin{equation}
    \left(\widetilde{\varphi}_a\right)^i = \varepsilon^{ij} \left(\varphi^*_a\right)_j\,, \qquad \text{with} \qquad \varepsilon = i \sigma_2\,,
\end{equation}
where the indices $i$ and $j$ refer to the weak-isospin component of the doublets $\varphi_a$ ($a=1,2$). 
We note that the position of the family index $a$ is relevant, since $\varphi_a$ transforms under the fundamental representation of $U(2)_H$ (the group describing unitary mixing of the two doublets), while $\varphi^*_a$ and $\widetilde{\varphi}_a$ transform under the anti-fundamental representation of the family symmetry group.\\

For each fermion species $f = u,d,e$, the Yukawa matrices $y_f$ and $\epsilon_f$ can in fact be unified into a single object $\mathcal{F} = \mathcal{U}, \mathcal{D}, \mathcal{E}$, transforming under the (anti-)fundamental representation of the family group. Namely, we define
\begin{subequations}
    \begin{align}
        \mathcal{U}^a = \begin{pmatrix} y_u \\  \epsilon_u \end{pmatrix}\ \ &, \ \ \mathcal{U}^\dagger_a = \begin{pmatrix} y_u^\dagger & \epsilon_u^\dagger \end{pmatrix}\,,\\ 
        \mathcal{D}_a = \begin{pmatrix} y_d & \epsilon_d \end{pmatrix}\ \ &, \ \ {\mathcal{D}^\dagger}^a = \begin{pmatrix} y_d^\dagger \\ \epsilon_d^\dagger \end{pmatrix}\,,\\
        \mathcal{E}_a = \begin{pmatrix} y_e & \epsilon_e \end{pmatrix}\ \ &, \ \ {\mathcal{E}^\dagger}^a = \begin{pmatrix} y_e^\dagger \\ \epsilon_e^\dagger \end{pmatrix}\,,
    \end{align}
\end{subequations}
in order to rewrite the Yukawa Lagrangian~\eqref{eq:yukLag} in the more compact form
\begin{align}
    -\mathcal{L}_Y = \Big[
    \overline{Q}_L\;\mathcal{U}^a\,\widetilde{\varphi}_a\; u_R
    + \overline{Q}_L\;\mathcal{D}_a\,\varphi^a\; d_R
    +  \overline{L}\;\mathcal{E}_a\,\varphi^a\;e_R
\Big] + \mathrm{h.c.}\,,
\end{align}
manifestly invariant under a change of basis in the Higgs family space, that is, a unitary transformation $U$ of the two doublets, provided
\begin{align}
    \varphi^a \rightarrow U^a_{\ \, b} \, \varphi^b \quad \Rightarrow \quad \mathcal{D}_a &\rightarrow \mathcal{D}_b \left(U^\dagger\right)^b_{\ a}, &  \mathcal{E}_a &\rightarrow \mathcal{E}_b \left(U^\dagger\right)^b_{\ a}, & \mathcal{U}^a &\rightarrow U^a_{\ \, b} \, \mathcal{U}^b\,.
\end{align}
While the Yukawa interactions themselves are only linear in the scalar fields (and the Yukawa matrices), some physically relevant quantities such as the fermion mass matrix squared, $M_F^2$, depend on them quadratically. For such quantities, the bilinear formalism can be appropriately used and can in fact be extended to the basis-dependent objects $\mathcal{D}$, $\mathcal{E}$ and $\mathcal{U}$. In analogy with the definition of the bilinear fields 
in~\eqref{eq:KbarToBilinears} which we repeat here for clarity,
\begin{equation}
    \twomat{K}^a_{\,\ b} = \varphi^a \varphi_b^\dagger = \frac{1}{2} K^\mu \left(\sigma_\mu\right)^a_{\,\ b}\,,
\end{equation}
we may write
\begin{equation} \label{eq:Yudef}
    \mathcal{U}^a \mathcal{U}_b^\dagger \equiv \frac{1}{2} Y_u^\mu \left(\sigma_\mu\right)^a_{\,\ b}\,,
\end{equation}
as the definition of the bilinear up-type Yukawa coupling $Y_u$. Similarly, the four-component bilinear Yukawa couplings $Y_d$ and $Y_e$ can be defined through
\begin{equation}  \label{eq:Ydedef}
    \mathcal{D}_a {\mathcal{D}^\dagger}^b = \frac{1}{2} Y_d^\mu \left(\sigma_\mu\right)^b_{\,\ a}\,, \qquad 
    \mathcal{E}_a {\mathcal{E}^\dagger}^b = \frac{1}{2} Y_e^\mu \left(\sigma_\mu\right)^b_{\,\ a}
\end{equation}
with
\begin{equation}
    \left(D_a (D^\dagger)^b\right)^{ik} = D_a^{ij} \left(D^*\right)^{b}_{kj} = \left(D^*\right)^{b}_{kj} D_a^{ij}\,.
\end{equation}

Contracting both sides of~\eqref{eq:Yudef} with $\left(\sigma^\nu\right)^b_{\,\ a}$, and both sides 
of~\eqref{eq:Ydedef} with $\left(\sigma^\nu\right)^a_{\,\ b}$ allows to obtain the expressions of $Y_{u,d,e}$:
\begin{equation}
    Y_u^\nu = \left(\sigma^\nu\right)^b_{\,\ a} \mathcal{U}^a \mathcal{U}_b^\dagger, \qquad Y_d^\nu = \left(\sigma^\nu\right)^a_{\,\ b} \mathcal{D}_a {\mathcal{D}^b}^\dagger, \quad , \qquad Y_e^\nu = \left(\sigma^\nu\right)^a_{\,\ b} \mathcal{E}_a {\mathcal{E}^b}^\dagger\,.
\end{equation}
Explicitly, the components of $Y_f$ ($f=u,d,e$) are given by the hermitian matrices
\begin{equation}
    Y_u = \begin{pmatrix}
        y_u y_u^\dagger + \epsilon_u \epsilon_u^\dagger\\
        y_u \epsilon_u^\dagger + \epsilon_u y_u^\dagger \\
        i\left(y_u \epsilon_u^\dagger - \epsilon_u y_u^\dagger\right)\\
        y_u y_u^\dagger - \epsilon_u \epsilon_u^\dagger
    \end{pmatrix}, \quad
    Y_d = \begin{pmatrix}
        y_d y_d^\dagger + \epsilon_d \epsilon_d^\dagger\\
        y_d \epsilon_d^\dagger + \epsilon_d y_d^\dagger \\
        -i\left(y_d \epsilon_d^\dagger - \epsilon_d y_d^\dagger\right)\\
        y_d y_d^\dagger - \epsilon_d \epsilon_d^\dagger
    \end{pmatrix}, \quad
    Y_e = \begin{pmatrix}
        y_e y_e^\dagger + \epsilon_e \epsilon_e^\dagger\\
        y_e \epsilon_e^\dagger + \epsilon_e y_e^\dagger \\
        -i\left(y_e \epsilon_e^\dagger - \epsilon_e y_e^\dagger\right)\\
        y_e y_e^\dagger - \epsilon_e \epsilon_e^\dagger
    \end{pmatrix}\,.
\end{equation}
Under changes of basis, the four-component bilinear Yukawa couplings transform in the same way as $\KT$, namely (for $f = u,d,e$)
\begin{equation}
    Y_f^0 \rightarrow Y_f^0, \qquad Y_f^a \rightarrow R(U)^{ab} \,  Y_f^b, \ a=1,2,3
\end{equation}
where $R(U)$ is an orthogonal $3\times3$ matrix defined in Eq.~\eqref{eq:bilinearRtransfo}.\\

From these transformation properties, we may define in particular the Yukawa couplings $\bar{\mathcal{U}}$, $\bar{\mathcal{D}}$ and $\bar{\mathcal{E}}$ obtained after performing the change of basis which diagonalises the neutral mass matrix, described in Eqs.~\eqref{eq:diagonalBasisNeutral} and \eqref{bilbasis}. Similarly, the four-component bilinear Yukawa couplings in such a basis are denoted
\begin{equation}
    \bar{Y}_f = \begin{pmatrix}
        Y_f^0 \\
        \bar{\tvec{Y}}_f
    \end{pmatrix}
\end{equation}
for $f=u,d,e$.

\subsection{Yukawa couplings in the mass basis}

Having established a basis-independent formalism to describe the Yukawa sector of any THDM, we are now interested in computing the tree-level Yukawa vertices involving the physical states, \textit{i.e.}~in expressing such vertices in the mass basis. Having determined the form of the scalar generators in such a basis in Sec.~\ref{sec:scalar-gen}, one way to proceed to compute the Yukawa vertices 
is to rely on the gauge-invariance of the Yukawa interaction Lagrangian. First, let us note that the Yukawa interactions can generically be written as
\begin{equation} \label{eq:genericYukLag}
    -\mathcal{L}_Y = \frac{1}{2} y^{IJ}_i \phi^i \psi^I \psi^J + \mathrm{h.c.}
\end{equation}
where all left-handed two-component spinors of the model were gathered in the fermion multiplet $\psi$. Under $\smgroup$, the scalar and fermion multiplets simultaneously transform as
\begin{equation}
    \phi^i \rightarrow \left(\unitmatrix + i \theta^A T^\phi_A\right)^{ij} \phi^j, \qquad \psi^{I} \rightarrow \left(\unitmatrix + i \theta^A T^\psi_A\right)^{IJ} \psi^{J}\,, \qquad A=1,\dots,12\,,
\end{equation}
with $T^\psi$ the gauge generators of the left-handed fermion multiplet. Extending the notations of the previous section, the generators $T^{\phi, \psi}_1$,  $T^{\phi, \psi}_{2,3,4}$ and $T^{\phi, \psi}_{5,\dots,12}$  correspond to $U(1)_Y$, $\mathrm{SU}(2)_L$, and $\mathrm{SU}(3)_C$ transformations respectively. The invariance of \eqref{eq:genericYukLag} under infinitesimal gauge transformations implies\footnote{Note that the generators $T^\phi_A$ are purely imaginary such that ${T^\phi_A}^* = -T^\phi_A$.}
\begin{equation}
    \left({T^\psi_A}^*\right)^{IK} y_i^{KJ} + y_i^{IK} \left(T^\psi_A\right)^{KJ} - \left(T^\phi_A\right)^{ij} y_j^{IJ} = 0
\end{equation}
for all values of the indices $A, I, J$ and $i$. Since this relation must hold independently of the basis chosen to express the various types of fields, we may in particular express it in the scalar mass basis, thus obtaining
\begin{equation}
    \left({T^\psi_A}^*\right)^{IK} \bar{y}_i^{KJ} + \bar{y}_i^{IK} \left(T^\psi_A\right)^{KJ} - \left(\bar{T}^\phi_A\right)^{ij} \bar{y}_j^{IJ} = 0\,,
\end{equation}
with $\bar{y}$ denoting the tensor of Yukawa couplings obtained after rotation of its scalar index towards the mass basis. This relation is linear in $\bar{y}$ and can be solved without too much difficulty, based on the known expressions of $T^\psi$ and $\bar{T}^\phi$. While the linear system of equations thus obtained is underdetermined, we may, as previously done in the gauge sector, supplement it with a set of relations which are independent on the basis chosen to express the scalar fields. For instance, imposing
\begin{equation}
    \Gamma^\mu_i \Gamma^\nu_j y_i^{IJ} y_j^{JI} = \bar{\Gamma}^\mu_i \bar{\Gamma}^\nu_j \bar{y}_i^{IJ} \bar{y}_j^{JI}
\end{equation}
suffices to compute the entries of $\bar{y}_i^{IJ}$ unambiguously, completing the determination of the Yukawa interactions in the mass basis.\\

After symmetry breaking towards $\emgroup$, the Yukawa interaction Lagrangian can be expanded, in terms of two-component fermions, as
\begin{align}
    - \mathcal{L}_Y ={}& \bigg\{ h^a \left[\bar{\xi}^u_a \, u_L^\dagger u_R + \bar{\xi}^d_a \,  d_L^\dagger d_R + \bar{\xi}_a^e \, e_L^\dagger e_R\right] + G^0 \left[\bar{\xi}^u_{G^0} \, u_L^\dagger u_R + \bar{\xi}^d_{G^0} \, d_L^\dagger d_R + \bar{\xi}^e_{G^0} \, e_L^\dagger e_R\right] \nonumber\\
    &+ H^- \left[\bar{\xi}^{ud}_{H^-} \, u_L^\dagger d_R + \bar{\xi}^{\nu e}_{H^-} \, \nu_L^\dagger e_R \right] + G^- \left[\bar{\xi}^{ud}_{G^-} \, u_L^\dagger d_R + \bar{\xi}^{\nu e}_{G^-} \, \nu_L^\dagger e_R \right] \label{eq:expandedYukLag}\\
    &+ H^+ \, \bar{\xi}^{ud}_{H^+} \, d_L^\dagger u_R + G^+ \, \bar{\xi}^{ud}_{G^+} \, d_L^\dagger u_R \bigg\} + \mathrm{h.c.}\, , \nonumber
\end{align}
where a sum on the index $a=1,2,3$ (\textit{i.e.}~on the three neutral scalars) is implied in the first term. In addition, if more than one generation of fermions are considered, we assume that the fermion mass matrices have been properly (bi-)diagonalised so that $u_{L,R}$, $d_{L,R}$ and $e_{L,R}$ correspond to fermion mass eigenstates. The expressions of all Yukawa matrices 
involved in~\eqref{eq:expandedYukLag} in terms of bilinear fields are given in App.~\ref{app:yukawaCouplings}. These expressions involve the complex two-vector $\bar{\kappa}^a$, defined such that
\begin{equation} \label{eq:kappaDefinition}
    \twomat{\bar{K}}^a_{\ \, b} = \frac{1}{2} \bar{K}^\mu \left(\sigma_\mu\right)^a_{\ \, b} \equiv \bar{\kappa}^a \bar{\kappa}^*_b\,.
\end{equation}
It should be noted that such a decomposition is made possible by $\twomat{K}$ being of rank 1 at a charge-conserving minimum (and, more generally, within the charge-conserving hypersurface). Explicitly, $\bar{\kappa}$ can be written
\begin{equation} \label{eq:kappaExpression}
    \bar{\kappa} = \sqrt{\frac{K_0}{2}} \frac{1}{\sqrt{1 + \bar{k}_3}}\begin{pmatrix}
        1 + \bar{k}_3 \\
        \bar{k}_1 + i \bar{k}_2
    \end{pmatrix} = \sqrt{\frac{K_0}{2}} \begin{pmatrix}
        \sqrt{1 + \bar{k}_3} \\
        \sqrt{1 - \bar{k}_3} \, e^{i \zeta}
    \end{pmatrix}\,,
\end{equation}
where the phase $\zeta$ was defined such that
\begin{equation}
    \bar{k}_1 + i \bar{k}_2 = \sqrt{\bar{k}_1^2 + \bar{k}_2^2} \, e^{i \zeta} = \sqrt{1 - \bar{k}_3^2} \, e^{i \zeta}\,.
\end{equation}
From~\eqref{eq:kappaExpression}, it is straightforward to check that
\begin{equation}
    \bar{\kappa}^a \bar{\kappa}^*_b = \frac{K_0}{2} \begin{pmatrix}
        1 + \bar{k}_3 & \bar{k}_1 - i \bar{k}_2 \\
        \bar{k}_1 + i \bar{k}_2 & 1 - \bar{k}_3
    \end{pmatrix}
\end{equation}
as expected from~\eqref{eq:kappaDefinition}. The complex 2-vector $\bar{\kappa}$ thus defined allows to write the various Yukawa vertices in a very compact form. For instance, the $h^a u_L^\dagger u_R$ and $h^a d_L^\dagger d_R$ interactions given in~\eqref{eq:hupYuk} and~\eqref{eq:hdownYuk} read
\begin{equation}
    \bar{\xi}^u_a = \frac{1}{\sqrt{2 K_0}} \, \left(\sigma_a\right)^\alpha_{\ \, \beta} \bar{\kappa}^*_\alpha \bar{\mathcal{U}}^\beta\,,
    \qquad \bar{\xi}^d_a = \frac{1}{\sqrt{2 K_0}} \, \left(\sigma_a\right)^\alpha_{\ \, \beta} \bar{\mathcal{D}}_\alpha \bar{\kappa}^\beta\,.
\end{equation}

\subsection{Fermion masses}

After symmetry breaking, the scalar multiplet acquires a vacuum expectation value according to~\eqref{eq:phiVacuum}, and the part of the Lagrangian describing the fermion mass terms reads
\begin{equation}
    -\mathcal{L} \supset \left\{ u_L^\dagger M_u u_R + d_L^\dagger M_d d_R + e_L^\dagger M_e e_R \right\} + \mathrm{h.c.}\,,
\end{equation}
where, for $f=u,d,e$,
\begin{equation}
    M_f = \sqrt{2 K_0} \, \bar{k}^a \bar{\xi}_a^f  = \sqrt{2 K_0} \, k^a \xi_a^f\,.
\end{equation}
For instance, we find for the up-type quarks
\begin{equation}
    M_u = \sqrt{2 K_0} \, k^a \xi^u_a = \left(k^a \sigma_a\right)^\alpha_{\,\ \beta} \kappa^*_\alpha \mathcal{U}^\beta\,.
\end{equation}
To further simplify this expression, we note that since
\begin{equation}
    \twomat{K}^\alpha_{\ \, \beta} = \frac{1}{2} K^\mu \left(\sigma_\mu\right)^\alpha_{\,\ \beta} = \frac{K_0}{2} \left[ \delta^\alpha_{\,\ \beta} + \left(k^a \sigma_a\right)^\alpha_{\,\ \beta}\right] = \kappa^\alpha \kappa^*_\beta\,,
\end{equation}
we find
\begin{equation}
    \left(k^a \sigma_a\right)^\alpha_{\,\ \beta} = \frac{2}{K_0} \kappa^\alpha \kappa^*_\beta - \delta^\alpha_{\,\ \beta}\,,
\end{equation}
so that
\begin{equation}
    M_u = \left[ \frac{2}{K_0} \kappa^\alpha \kappa^*_\beta - \delta^\alpha_{\,\ \beta} \right]\kappa^*_\alpha \mathcal{U}^\beta = \left(\frac{2}{K_0} \left|\kappa\right|^2 - 1\right) \kappa^*_\alpha \mathcal{U}^\alpha \,.
\end{equation}
Finally,
\begin{equation}
    \left|\kappa\right|^2 = \kappa^*_\alpha \kappa^\alpha = \tr \uline K = K_0
\end{equation}
so $M_u$ is simply given by
\begin{equation}
    M_u = \kappa^*_\alpha \mathcal{U}^\alpha\,.
\end{equation}
In turn, the up-type quark mass matrix squared, $M_u^2$ is computed as
\begin{equation}
    M_u^2 = M_u M_u^\dagger = \kappa^*_\alpha \kappa^\beta \mathcal{U}^\alpha \mathcal{U}^\dagger_\beta = \twomat{K}^\beta_{\,\ \alpha} \mathcal{U}^\alpha \mathcal{U}^\dagger_\beta = \frac{1}{2} K^\mu \left(\sigma_\mu\right)^\beta_{\,\ \alpha}  \mathcal{U}^\alpha \mathcal{U}^\dagger_\beta = \frac{1}{2} K_\mu Y_u^\mu \, .
\end{equation}
Similarly, we obtain for the down-type quark and lepton mass matrices:
\begin{align}
    && && M_d &= \bar{\mathcal{D}}_\alpha \bar{\kappa}^\alpha, & M_d^2 &= \frac{1}{2} K_\mu Y_d^\mu\,, && &&\\
    && && M_e &= \bar{\mathcal{E}}_\alpha \bar{\kappa}^\alpha, & M_e^2 &= \frac{1}{2} K_\mu Y_e^\mu\,. && &&
\end{align}


\section{Conclusions}
\label{sec:conclusions}

In the THDM potential it has been shown that gauge-invariant expressions give new insights~\cite{Velhinho:1994np}. 
With the introduction of bilinears~\cite{Nagel:2004sw,Nishi:2006tg,Maniatis:2006fs}
gauge-invariants have been introduced which have a one-to-one correspondence to the Higgs-boson doublets
 - except for the non-physical gauge redundancies. 
 In particular, 
it has been shown, that bilinears form Minkowski-type four vectors. Stability, electroweak symmetry breaking and the 
symmetries of the potential can be studied in a concise form based on bilinears~\cite{Maniatis:2006fs,Maniatis:2011qu,Ivanov:2007de}.
 
Here we have extended this gauge-invariant bilinear formalism to the squared mass matrices and the interaction terms of the THDM, that is, the trilinear and quartic scalar couplings, the gauge and Higgs boson interactions, and the Yukawa couplings. 

With the help of the connection $\Gamma$, \eqref{eq:defgamma}, between the component fields of the two doublets and the bilinears we were able to get a completely gauge-invariant form of the scalar squared mass matrix. We revealed the general expressions for the cases of full electroweak symmetry breaking as well as for the charge-conserving breaking case, without specifying the actual form of the potential. 
We derived the scalar mass spectrum for the general case in terms of gauge-invariant expressions. 
In particular the respective Goldstone modes appeared in a very transparent way.
From the gauge-invariant form of the mass matrix we then derived the cubic and quartic scalar interactions. 
Then, from the expressions of the scalar gauge generators in the mass-basis, we calculated the full set of vector-scalar couplings in a gauge-invariant form.
Finally, we have extended the bilinear formalism to the Yukawa sector and obtained expressions for the fermion mass matrices squared that are invariant under mixing of the two Higgs doublets. Starting again from the expression of the scalar generators and employing the gauge-invariance of the Yukawa Lagrangian, we obtained gauge-invariant expressions for the Yukawa couplings after electroweak symmetry breaking.\\

With these results, we have extended the bilinear formalism to all interactions in a general THDM, making it a truly powerful tool to study any specific THDM.



\appendix


\section{Structure of the scalar mass matrix}
\label{app:scalarMassMatrixStructure}

This appendix extends the discussion of section~\ref{sec:gaugeInvariantFormalism} on the structure of the scalar mass matrix. In particular, we provide explicit forms for the various quantities involved in the expression of the mass matrix,
at both, at a charge-breaking minimum and at a charge-conserving minimum. The various quantities expressed in this appendix will allow in turn to compute analytically the derivatives of scalar eigenvalues, as required to express the derivatives of the effective potential. First, let us recall that the general form of the scalar mass matrix in our formalism:
\begin{equation}\label{eq:massMatCano}
    \widehat{M}_s^2 = \widehat{\Delta}^\mu \partial_\mu V + \widehat{\Gamma} \mathcal{M} \widehat{\Gamma}^\trans \,.
\end{equation}
In particular, the hatted quantities correspond to
the canonical bases; see section~\ref{sec:gaugeInvariantFormalism}. In this basis,
the connection matrices $\Gamma$ are of the 
form~\eqref{eq:canonicalGamma}.
To simplify the forthcoming expressions, we define
\begin{equation}
    \alpha = \left(\frac{\KT^\trans \gT \KT}{K_0^2}\right)^{1/2} = \sqrt{1 - \tvec{k}^\trans \tvec{k}}
\end{equation}
so that $\gamma$~\eqref{eq:gamma} can be written
\begin{equation}
    \gamma = \sqrt{2 K_0} \begin{pmatrix}
       \alpha & \tvec{0}^\trans \\
       \tvec{k} & \unitmatrix_3
    \end{pmatrix}\,.
\end{equation}

We now want to derive the explicit form of 
 $\widehat{M}_s^2$ in a canonical basis. 
 We first need to examine the form taken by the $\widehat{\Delta}^\mu$ matrices. 
We may decompose the four symmetric $8 \times 8$ matrices $\widehat{\Delta}^\mu$ into $4 \times 4$ blocks
\begin{equation} \label{eq:Deltahat}
    \widehat{\Delta}^\mu = \begin{pmatrix}
        A_{44}^\mu & C_{44}^\mu \\[.15cm]
        \left(C_{44}^\mu\right)^\trans & B_{44}^\mu \\
    \end{pmatrix}\,.
\end{equation}
As we will show, $A_{44}^\mu$ and $C_{44}^\mu$ depend on the chosen basis, while $B_{44}^\mu$ is fixed by the requirement of working in a canonical basis. In fact, it is possible to determine $B_{44}^\mu$ by first computing the quantity
\begin{equation} \label{eq:defsig}
    \Sigma^{\mu\rho\sigma} = \Gamma^\rho_i \Delta^\mu_{ij} \Gamma^\sigma_j = \Gamma^\rho_i \Delta^\mu_{ji} \Gamma^\sigma_j
    = \Sigma^{\mu\sigma\rho}\,.
\end{equation}
or, in matrix notation
$\Sigma^\mu = \Gamma^\trans \Delta^\mu \Gamma$.
This quantity is constructed in a way that all field component indices are contracted and therefore it is gauge invariant. In order to get an explicit expression for $\Sigma^\mu$ we first compute
\begin{align} 
    \Gamma^\mu_i \Delta^\nu_{ia}  &= \Delta^\mu_{ij} \Delta^\nu_{ia} \phi^j \nonumber\\
    &= \partial_a \left( \Delta^\mu_{ij} \Delta^\nu_{ik} \phi^j \phi^k \right) -  \Delta^\mu_{ia} \Delta^\nu_{ik} \phi^k \nonumber\\
    &= \partial_a \left(\Gamma^2\right)^{\mu\nu} -  \Delta^\mu_{ia} \Delta^\nu_{ij} \phi^j \nonumber\\
    &= \Gamma^\lambda_a T^{\mu\nu}_\lambda -  \Gamma^\nu_{i} \Delta^\mu_{ia} \label{eq:deltaGammaRel} \,,
\end{align}
where the rank-3 symmetric tensor $T$ has been defined in~\eqref{eq:ga22}
and we have used~\eqref{eq:KphiDef}. By repeatedly applying~\eqref{eq:deltaGammaRel} in the definition of $\Sigma^\mu$, recalling that $\Gamma^2\equiv \Gamma^\trans \Gamma$, that is,
$(\Gamma^2)^{\rho \sigma} = \Gamma^\rho_i \Gamma^\sigma_i$,
we arrive at
\begin{equation} \label{eq:sigma}
    \Sigma^\mu = \frac{1}{2} \left[ \Gamma^2 T^\mu + T^\mu \Gamma^2 - \left(\Gamma^2\right)^{\mu\nu} T^\nu \right]\,.
\end{equation}
Let us show~\eqref{eq:sigma} in detail, starting from the definition
of $\Sigma^\mu$, \eqref{eq:defsig}, writing all indices explicitly
and using~\eqref{eq:deltaGammaRel} twice,
\begin{equation}
\Sigma^{\mu \rho \sigma} = 
\Gamma^\rho_i \Delta_{ij}^\mu \Gamma_j^\sigma
=
\Gamma_j^\lambda T^{\rho \mu}_\lambda \Gamma_j^\sigma 
-
\Gamma_i^\mu \Delta_{ij}^\rho \Gamma_j^\sigma
=
\Gamma_j^\lambda T^{\rho \mu}_\lambda \Gamma_j^\sigma
-
\Gamma_i^\mu ( \Gamma_i^\lambda T^{\rho \sigma}_\lambda - \Gamma_j^\rho \Delta_{ij}^\sigma)\,.
\end{equation}
For the first term on the right-hand side we get 
$\Gamma_j^\lambda T^{\rho \mu}_\lambda \Gamma_j^\sigma = T^\mu (\Gamma^2)^{\rho \sigma}$
and for the second term 
$\Gamma_i^\mu \Gamma_i^\lambda T^{\rho \sigma}_\lambda =
(\Gamma^2)^{\mu \lambda} T_\lambda^{\rho \sigma}$, for the third term
$\Gamma_i^\mu \Gamma_j^\rho \Delta_{ij}^\sigma =
\Gamma_i^\mu \Delta_{ij}^\sigma \Gamma_j^\rho =
\Gamma_j^\lambda T^{\mu \sigma}_\lambda \Gamma_j^\rho -
\Gamma_i^\sigma \Delta_{ij}^\mu \Gamma_j^\rho =
(\Gamma^2)^{\rho \sigma} T^\mu - 
\Sigma^{\mu \rho \sigma}$
altogether,
\begin{equation}
\Sigma^{\mu \rho \sigma} = 
T^\mu (\Gamma^2)^{\rho \sigma} -  (\Gamma^2)^{\mu \lambda} T_\lambda^{\rho \sigma} + (\Gamma^2)^{\rho \sigma} T^\mu - \Sigma^{\mu \rho \sigma} \,.
\end{equation}
This is equivalent to~\eqref{eq:sigma}. 

With the explicit form for the $T_\mu$ matrices 
given in~\eqref{eq:Texp},
\begin{equation}
    T_0 = 2 \unitmatrix_4, \quad T_i = 2 \begin{pmatrix}
    0 & \tvec{e}_i^\trans \\
    \tvec{e}_i & 0_{3\times3}
    \end{pmatrix}\,,
\end{equation}
we find
\begin{equation} \label{eq:sigmaExpr}
    \Sigma^0 = \Gamma^2 = 2\begin{pmatrix}
        K_0 & \tvec{K}^\trans \\
        \tvec{K} & K_0 \unitmatrix_3
    \end{pmatrix}, \quad
    \Sigma^i = 2 K_i \gT + 2 \begin{pmatrix}
        0 & K_0 \tvec{e}_i^\trans \\
        K_0 \tvec{e}_i & \tvec{e}_i \tvec{K}^\trans + \tvec{K} \tvec{e}_i^\trans \\
    \end{pmatrix}\,.
\end{equation}
On the other hand, we can express $\Sigma^\mu$,
\eqref{eq:defsig} in any basis, since it is gauge-invariantly defined. In the canonical basis we have 
\begin{equation}
    \Sigma^\mu = \Gamma^\trans \Delta^\mu \Gamma = \widehat{\Gamma}^\trans \widehat{\Delta}^\mu \widehat{\Gamma} = \begin{pmatrix}
    0_{4\times4} & \gamma^\trans
    \end{pmatrix}
    \begin{pmatrix}
        A_{44}^\mu & C_{44}^\mu \\[.15cm]
        \left(C_{44}^\trans\right)^\mu & B_{44}^\mu \\
    \end{pmatrix}
    \begin{pmatrix}
    0_{4\times4} \\ \gamma
    \end{pmatrix} = \gamma^\trans B_{44}^\mu \gamma\,.
\end{equation}
At a point of the bilinear field space where $\gamma$ is non-singular (i.e., where $\det(\gamma)=4 K_0^2 \alpha \ne 0$), 
the relation can be inverted using
\begin{equation}
    \gamma^{-1} = \frac{1}{\alpha \sqrt{2 K_0}} \begin{pmatrix}
    1& \tvec{0}^\trans \\
    - \tvec{k} &  \alpha \unitmatrix_3
    \end{pmatrix}
\end{equation}
and we finally get with~\eqref{eq:sigmaExpr}:
\begin{equation}\label{eq:Bdelta}
    B_{44}^\mu = \left(\gamma^\trans\right)^{-1} \Sigma^\mu \gamma^{-1} \quad  \text{that is} \quad B_{44}^0 = \unitmatrix_4, \quad B_{44}^i = - k_i \unitmatrix_4 + \begin{pmatrix}
        0 & \alpha \tvec{e}_i^\trans \\
        \alpha \tvec{e}_i & \tvec{e}_i \tvec{k}^\trans + \tvec{k} \tvec{e}_i^\trans
    \end{pmatrix}\,.
\end{equation}
Having found the result for $B_{44}^\mu$ valid in any canonical basis, we want to determine a possible form for $A_{44}^\mu$ and $C_{44}^\mu$. As stated above, these two matrices are not uniquely determined since a continuous set of canonical bases exist. However, requiring that $\widehat{\Delta}^\mu$ satisfies the same anti-commutation properties as $\Delta^\mu$ in \eqref{eq:anti-comm}, namely
\begin{equation}
    \left\{\widehat{\Delta}^\mu, \widehat{\Delta}^\nu\right\} = T^{\mu\nu}_\lambda \widehat{\Delta}^\lambda\,,
\end{equation}
and that the components $\widehat{\Delta}^a$, $a=1,2,3$ properly transform under rotations in the bilinear field space, allows us to first determine the set of possible expressions for $C_{44}^\mu$, and then for $A_{44}^\mu$. These expressions are valid anywhere in the field space, hence, in particular at the charge-conserving and the charge-breaking hypersurfaces. We may now provide a complete set of $A_{44}^\mu$, $B_{44}^\mu$ and $C_{44}^\mu$ matrices valid for any values of the bilinear fields but assuming a canonical basis:
\begin{align}
    A_{44}^0 &= \unitmatrix_4, & A_{44}^a &= k_a \unitmatrix_4 - \frac{1}{|\tvec{k}|^2} \begin{pmatrix}
        0 & \alpha k_a \tvec{k}^\trans \\
        \alpha k_a \tvec{k} \quad & 2 k_a \tvec{k} \tvec{k}^\trans\!\! -\!\! \left(\left[\tvec{e}_a\times\tvec{k}\right] \tvec{k}^\trans + \tvec{k} \left[\tvec{e}_a\times\tvec{k}\right]^\trans\right)
    \end{pmatrix}\label{eq:A44}\\[.2cm] 
    B_{44}^0 &= \unitmatrix_4, & B_{44}^a &= - k_a \unitmatrix_4 + \begin{pmatrix}
        0 & \alpha\, \tvec{e}_a^\trans \\
        \alpha\, \tvec{e}_a \quad & \tvec{e}_a \tvec{k}^\trans + \tvec{k} \tvec{e}_a^\trans
    \end{pmatrix}\,,
    \label{eq:B44}\\[.2cm]
    C_{44}^0 &= 0_{4\times4}, & C_{44}^a &= \frac{1}{|\tvec{k}|} \begin{pmatrix}
        0 & \left[\tvec{e}_a \times \tvec{k}\right]^\trans \\[.15cm]
        |\tvec{k}|^2 \tvec{e}_a - k_a \tvec{k} \quad 
        & \alpha \left(k_a \unitmatrix_3 - \tvec{e}_a \tvec{k}^\trans\right)
    \end{pmatrix}\label{eq:C44}\,.
\end{align}
For the case of a charge-conserving minimum we have $|\tvec{k}|=1$, corresponding to $\alpha=0$. 
In this case (see \eqref{eq:canonicalGamma})
\begin{equation} 
    \widehat{\Gamma} 
    = 
    \begin{pmatrix}
    0_{4\times4} \\ \gamma
    \end{pmatrix}
    =
    \begin{pmatrix}
    0_{4\times4} \\ 
     0_{1\times4} \\
       \gamma_3
    \end{pmatrix}
\end{equation}
since
\begin{equation}
    \gamma = \sqrt{2 K_0} \begin{pmatrix}
       \alpha & \tvec{0}^\trans \\
       \tvec{k} & \unitmatrix_3
    \end{pmatrix} \stackrel{CC}{=} \sqrt{2 K_0} \begin{pmatrix}
       0 & \tvec{0}^\trans \\
       \tvec{k} & \unitmatrix_3
    \end{pmatrix} 
    \equiv \begin{pmatrix}
        0_{1\times4} \\
       \gamma_3
    \end{pmatrix}\,.
\end{equation}
Here we have defined the $3 \times 4$ matrix~$\gamma_3$
with index~3 in order to distinguish it from the $4 \times 4$ matrix~$\gamma$.
As a consequence, the term $\widehat{\Gamma} \mathcal{M} \widehat{\Gamma}^\trans$ in the mass matrix~\eqref{eq:massMatCano} 
takes the form
\begin{equation} \label{eq:GMG}
    \widehat{\Gamma} \mathcal{M} \widehat{\Gamma}^\trans = \begin{pmatrix}
        0_{5\times5} & 0_{5\times3} \\
        0_{3\times5} & \gamma_3 \mathcal{M} \gamma_3^\trans
    \end{pmatrix}\, .
\end{equation}

Hence, at a charge-conserving minimum, the block-diagonal structure of $\widehat{\Delta}^\mu$ is more appropriately reorganised as
\begin{equation}
    \widehat{\Delta}^\mu = \begin{pmatrix}
        A_{55}^\mu & C_{53}^\mu \\[.15cm]
        C_{35}^\mu & B_{33}^\mu
    \end{pmatrix}\,,
\end{equation}
where the expression of the newly defined blocks $A_{55}^\mu$, $B_{33}^\mu$ and $C_{53}^\mu$ can be readily inferred from Eqs.~\eqref{eq:A44}--\eqref{eq:C44}. Next, we express the first term of~\eqref{eq:massMatCano} in the form
\begin{equation}
     \widehat{\Delta}^\mu \partial_\mu V \stackrel{CC}{=} 2u \, \big(\gT \KT\big)_\mu \widehat{\Delta}^\mu = 2u \begin{pmatrix}
        A_{55} & C_{53} \\[.15cm]
        C_{35} & B_{33}
    \end{pmatrix}\,,
\end{equation}
with
\begin{multline}
    A_{55} = 2 K_0 \begin{pmatrix}
        0 & \tvec{0}^\trans & 0 \\
        \tvec{0} & \tvec{k} \tvec{k}^\trans & \tvec{0}\\
        0 & \tvec{0}^\trans & 1\\
    \end{pmatrix},\quad 
    B_{33} = 2 K_0 \left(\unitmatrix_3 - \tvec{k} \tvec{k}^\trans\right) = - \gamma_3 \gT \gamma_3^\trans,\\
    C_{53} = \alpha K_0 \begin{pmatrix}
        \tvec{0}^\trans \\
        \tvec{k} \tvec{k}^\trans - \unitmatrix_3\\
        - \tvec{k^\trans}\\
    \end{pmatrix} = 0\,.
\end{multline}
Note that although $\alpha$ and therefore $C_{53}$ vanish at a charge-conserving minimum, for later convenience we want to have the expression for $\widehat{\Delta}^\mu$ not only at the minimum of the potential. 
Eventually, we have found that the off-diagonal blocks of $\widehat{M}_s^2$ of~\eqref{eq:massMatCano} vanish,
simultaneously for the term
$\widehat{\Delta}^\mu \partial_\mu V$ and for the term $\widehat{\Gamma} \mathcal{M} \widehat{\Gamma}^\trans$.
This shows the convenience to choose a canonical basis,
where this simple structure of the scalar mass matrix
appears. We may thus finally write with~\eqref{eq:GMG},
\begin{equation}
    \widehat{M}_s^2 \stackrel{CC}{=} 2u \, \big(\gT \KT\big)_\mu \widehat{\Delta}^\mu + \widehat{\Gamma} \mathcal{M} \widehat{\Gamma}^\trans = \begin{pmatrix}
        \widehat{\mathcal{M}}^2_\mathrm{CC} & 0_{5\times3}\\
        0_{3\times5} & \widehat{\mathcal{M}}^2_\mathrm{neutral}
    \end{pmatrix}\,.
\end{equation}
The block
\begin{equation}
    \widehat{\mathcal{M}}^2_\mathrm{CC} = 2u A_{55}
\end{equation}
fully contains the Goldstone sector as well as the two massive charged Higgs fields~$H^\pm$. We can diagonalise this matrix
\begin{equation}
    \bar{\mathcal{M}}^2_\mathrm{CC} = U_\mathrm{CC} \widehat{\mathcal{M}}^2_\mathrm{CC} U_\mathrm{CC}^\trans = \diag \begin{pmatrix} 0, & 0, & 0, & 4u K_0, & 4u K_0\end{pmatrix} = \diag \begin{pmatrix} 0, & 0, & 0, & m_{H^\pm}^2, & m_{H^\pm}^2 \end{pmatrix}\,.
\end{equation}
The rotation matrix $U_\mathrm{CC}$ can explicitly be decomposed as
\begin{equation}
    U_\mathrm{CC} = \begin{pmatrix}
        1 & \tvec{0}^\trans & 0\\
        \tvec{0} & R_H & \tvec{0}\\
        0 & \tvec{0}^\trans & 1
    \end{pmatrix}
\end{equation}
with $R_H$ given by the $3\times3$ matrix in the bilinear field space which rotates~$\tvec{k}$ to the Higgs basis with $\tvec{k} = (0, 0, 1)^\trans$. Turning to the neutral sector, the matrix
\begin{equation} \label{eq:Mh0}
    \widehat{\mathcal{M}}^2_\mathrm{neutral} = \gamma_3 \left(\mathcal{M} - 2 u \gT\right) \gamma_3^\trans
\end{equation}
contains the masses of the three neutral states $h$, $H$ and $A$. 
 It is remarkable that $\widehat{\mathcal{M}}^2_\mathrm{neutral}$ can in fact always (that is, not only for the tree-level case) be diagonalised by a rotation in the bilinear field space, \textit{i.e.}~a change of basis:
\begin{equation}
    \bar{\mathcal{M}}_\mathrm{neutral} = R\widehat{\mathcal{M}}^2_\mathrm{neutral}R^\trans = \diag\begin{pmatrix}m_1^2, & m_2^2, & m_3^2\end{pmatrix}\,.
\end{equation}
The orthogonal rotation which diagonalises the $8\times 8$ scalar mass matrix in the canonical basis introduced above reads
\begin{equation}\label{eq:Ubar}
    \bar{U} = \begin{pmatrix}
        U_{CC} & 0_{5 \times 3} \\
        0_{3 \times 5} & R
    \end{pmatrix}.
\end{equation}
With this rotation matrix $\bar{U}$ we get the transformations of $\widehat{\Delta}$, $\widehat{\Gamma}$ from the canonical basis to the
Higgs basis:
\begin{equation} \label{eq:DeltaGammaBar}
    \bar{\Delta}^\mu = \bar{U} \widehat{\Delta}^\mu \bar{U}^\trans, \qquad \bar{\Gamma} = \bar{U} \widehat{\Gamma}\,.
\end{equation}
Gathering up the above results, the scalar mass matrix in the diagonal basis reads
\begin{equation}
    \bar{M}_s^2 = 2 u \gK_\mu \bar{\Delta}^\mu + \bar{\Gamma} \mathcal{M} \bar{\Gamma}^\trans = \diag\begin{pmatrix}0
    , & 0, & 0, & m_{H^\pm}^2, & m_{H^\pm}^2, & m_1^2, & m_2^2, & m_3^2\end{pmatrix}
\end{equation}
This concludes our discussion on the structure of the scalar mass matrix at a charge-conserving minimum. We stress that all of the above is valid at any order in perturbation theory. We think that this illustrates the convenience of working with the gauge-invariant formalism.

As an example the scalar mass matrix 
squared~\eqref{eq:Mh0} for the tree-level THDM potential
is computed in~\eqref{eq:MneutralTreeLevel}.


\section{THDM couplings}
\label{app:couplings}

In this section we present the analytic couplings of the general THDM in gauge invariant form. We recall that the indices $a, b, c, d, e \in \{1,2,3\}$ denote the three neutral Higgs-boson scalars. The squared masses of the three neutral scalars are denoted by $m_a^2$. For repeatedly appearing indices $a, b, c, d, e$ we employ as usual the sum convention. 
Note, that we do not distinguish between upper and lower indices.
We note that the index of the masses of the scalars does not imply any summation, for instance the expression $k^a m_a$ in~\eqref{eq:fa} is simply a vector with component~$a$, whereas in an expression like
$\varepsilon_{abc} \bar{k}^a m_a^2$ 
we have to sum over the index~$a$ because
it appears twice in tensor components, that is,
in $\varepsilon_{abc}$ and in $\bar{k}^a$.

\subsection{Scalar couplings}
\label{sub:scalarint}
The cubic and quartic neutral scalar couplings can be found in 
section~\ref{sec:treeLeveLScalarCouplings}. We repeat here for convenience the 3-vector $\bar{f}^a$ 
in terms of which some of the couplings below are expressed:
\begin{align} \label{eq:fa}
    \bar{f}^a &= 8 K_0 \left(\eta_{00} \bar{k}^a + \bar{\eta}^a\right) - \bar{k}^a m_a^2\,.
\end{align}
The scalar mass squared matrix~$\bar{\mathcal{M}}^2_{\text{neutral}}$ is 
given in~\eqref{eq:Mbar0tree}.
The 3-vectors $\tvec{x}$ and $\tvec{y}$ are defined in~\eqref{eq:xykRelations}.
The couplings then read:
\\

\noindent
{\bf $h_a$ - $h_b$ - $G^0$}
\begin{equation} \label{eq:abG}
    \bar{\lambda}_{abG^0} = \frac{1}{\sqrt{2 K_0}}
    \varepsilon_{abc} \bar{k}^c 
    \big(m_b^2-m_a^2 \big)
\end{equation}

\noindent
{\bf $h$ - $G^0$ - $G^0$}
\begin{equation}
    \bar{\lambda}_{aG^0G^0} = \frac{1}{\sqrt{2 K_0}}
    \bar{k}^a m_a^2
\end{equation}

\noindent
{\bf $h_a$ - $H^\pm$ - $H^\pm$}
\begin{equation}
    \bar{\lambda}_{aH^\pm H^\pm} = \frac{\bar{f}^a}{\sqrt{2 K_0}}
\end{equation}

\noindent
{\bf $h_a$ - $G^\pm$ - $G^\pm$}
\begin{equation}
    \bar{\lambda}_{aG^\pm G^\pm} = \frac{1}{\sqrt{2 K_0}} \bar{k}^a m_a^2
\end{equation}

\noindent
{\bf $h_a$ - $G^\pm$ - $H^\pm$}
\begin{equation}
    \bar{\lambda}_{aG^\pm H^\pm} = \frac{1}{\sqrt{2 K_0}} 
        \big(m_a^2 - m_{H^\pm}^2 \big)
        \big(
        x^a + i y^a
        \big)
\end{equation}

\noindent
{\bf $h_a$ - $h_b$ - $h_c$ - $h_d$}
\begin{equation}
    \bar{\lambda}_{abcd} =
    g_{abcd}+ g_{cdab} + g_{acbd} + g_{bdac} + g_{adbc} + g_{bcad}\,,
\end{equation}
with the auxiliary function
\begin{multline}
    g_{abcd}=
    \frac{\delta^{ab}}{2 K_0}
        \bigg(
        \frac{\delta^{cd}}{2}
        \bar{\tvec{k}}^\trans \bar{\mathcal M}^2_{\text neutral} \bar{\tvec{k}}
        +
         \bar{k}^c \bar{k}^d \big( m_a^2 + m_b^2 - m_c^2 - m_d^2 \big)
         \bigg)
         \\
        + 4 \big(
        \delta^{ab} - \bar{k}^a \bar{k}^b \big)
        \big( \bar{k}^c \bar{k}^d 
        (\eta_{00} - \frac{m_{H^\pm}^2}{4 K_0} )
        + \bar{\eta}^c \bar{k}^d + \bar{\eta}^d \bar{k}^c
        - \bar{\tvec{\eta}}^\trans \bar{\tvec{k}}
             \delta^{cd}
            \big)\, .
\end{multline}

\noindent
{\bf $h_a$ - $h_b$ - $h_c$ - $G^0$}
\begin{equation}
    \bar{\lambda}_{abcG^0} =
    f'_{abc}+
    f'_{bca}+
    f'_{cab}\, ,
\end{equation}
with the auxiliary function
\begin{equation}
    f'_{abc}=
    \frac{1}{2 K_0}
        \bigg(
        8 K_0
            \big(
            \delta^{ab}-
            \bar{k}^a \bar{k}^b 
            \big)
        (\bar{\tvec{\eta}} \times \bar{\tvec{k}})^c 
        - \frac{1}{2}
        \delta^{ab}
        \varepsilon^{cde}
        \big( m_d^2 -m_e^2 \big)
        \bar{k}^d \bar{k}^e
        -\varepsilon^{abd}
        \bar{k}^d
            \big(
            m_a^2 - m_b^2
            \big)
        \bar{k}^c
        \bigg)\, .
\end{equation}

\noindent
{\bf $h_a$ - $h_b$ - $G^0$ - $G^0$}
\begin{multline}
     \bar{\lambda}_{abG^0 G^0} = \frac{1}{2 K_0}
        \bigg(
        8 K_0 (\delta^{ab}-\bar{k}^a \bar{k}^b) 
            \big(
            \eta_{00} - \frac{m_{H^pm}^2}{4 K_0} + \bar{\tvec{\eta}}^\trans \bar{\tvec{k}}
            \big)\\
        + \bar{k}^a \bar{k}^b \big( m_a^2 + m_b^2 \big) - \delta^{ab}  
        \bar{\tvec{k}}^\trans \bar{\mathcal M}^2_{\text neutral} \bar{\tvec{k}}
          + 2 \varepsilon^{cad} \varepsilon^{cbe} \bar{k}^d \bar{k}^e m_c^2
        \bigg)
\end{multline}

\noindent
{\bf $h_a$ - $G^0$ - $G^0$ - $G^0$}
\begin{equation}
    \bar{\lambda}_{aG^0G^0 G^0} = \frac{3}{4 K_0}
        \varepsilon^{abc} \bar{k}^b \bar{k}^c (m_b^2 - m_c^2)
\end{equation}

\noindent
{\bf $G^0$ - $G^0$ - $G^0$ - $G^0$}
\begin{equation}
    \bar{\lambda}_{G^0G^0G^0G^0} = \frac{3}{2 K_0}
    \bar{\tvec{k}}^\trans \bar{\mathcal M}^2_{\text neutral} \bar{\tvec{k}} 
\end{equation}

\noindent
{\bf $h_a$ - $h_b$ - $H^\pm$ - $H^\pm$}
\begin{multline}
    \bar{\lambda}_{abH^\pm H^\pm} = \frac{1}{2 K_0}
        \bigg(
        \delta^{ab} 
                \big(  
            \bar{\tvec{k}}^\trans \bar{\mathcal M}^2_{\text neutral} \bar{\tvec{k}} - 16 K_0 \bar{\tvec{\eta}}^\trans \bar{\tvec{k}} 
                \big) 
            \\
        + \bar{k}^a \bar{k}^b \big ( 8 K_0 (\eta_{00} + 
            \bar{\tvec{\eta}}^\trans \bar{\tvec{k}}  ) - (m_a^2 + m_b^2) \big)
        + 8 K_0 \big( \bar{k}^a \bar{\eta}^b + \bar{k}^b \bar{\eta}^a \big)
        \bigg)
\end{multline}

\noindent
{\bf $h_a$ - $G^0$ - $H^\pm$ - $H^\pm$}
\begin{equation}
    \bar{\lambda}_{aG^0 H^\pm H^\pm} = \frac{1}{2 K_0}
        \varepsilon^{abc} \bar{f}^b \bar{k}^b \bar{k}^c
\end{equation}

\noindent
{\bf $G^0$ - $G^0$ - $H^\pm$ - $H^\pm$}
\begin{equation}
    \bar{\lambda}_{G^0 G^0 H^\pm H^\pm} = \frac{1}{2 K_0} \bar{k}^a \bar{f}^a
\end{equation}

\noindent
{\bf $h_a$ - $h_b$ - $G^\pm$ - $G^\pm$}
\begin{equation}
    \bar{\lambda}_{ab G^\pm G^\pm} = \frac{1}{2 K_0} 
        \bigg(
        8 K_0 
            (
            \delta^{ab} - \bar{k}^a \bar{k}^b
            ) ( \eta_{00} + \bar{\tvec{\eta}}^\trans \bar{\tvec{k}} )
            -\delta^{ab}  
            \bar{\tvec{k}}^\trans \bar{\mathcal M}^2_{\text neutral} \bar{\tvec{k}} + 
            \bar{k}^a \bar{k}^b (m_a^2 + m_b^2)
        \bigg)
\end{equation}

\noindent
{\bf $h_a$ - $G^0$ - $G^\pm$ - $G^\pm$}
\begin{equation}
    \bar{\lambda}_{a G^0 G^\pm G^\pm} = \frac{1}{4 K_0} 
        \varepsilon^{abc} \bar{k}^b \bar{k}^c (m_b^2 - m_c^2)
\end{equation}

\noindent
{\bf $G^0$ - $G^0$ - $G^\pm$ - $G^\pm$}
\begin{equation}
    \bar{\lambda}_{G^0 G^0 G^\pm G^\pm} = \frac{1}{2 K_0} 
    \bar{\tvec{k}}^\trans \bar{\mathcal M}^2_{\text neutral} \bar{\tvec{k}}
\end{equation}

\noindent
{\bf $h_a$ - $h_b$ - $G^\pm$ - $H^\pm$}
\begin{multline}
    \bar{\lambda}_{ab G^\pm H^\pm} = 
        4 \bar{\eta}^c (x^c + i y^c)
        (\delta^{ab} - \bar{k}^a \bar{k}^b )
        +\\
    \frac{1}{2 K_0} 
        \bigg(
        \bar{k}^a (x^b + i y^b) (m_b^2 - m_{H^\pm}^2)
        + \bar{k}^b (x^a + i y^a) (m_a^2 - m_{H^\pm}^2)
        - \bar{k}^c (x^c + i y^c) m_c^2
         \delta^{ab} 
         \bigg)
\end{multline}

\noindent
{\bf $h_a$ - $G^0$ - $G^\pm$ - $H^\pm$}\\
\begin{equation}
    \bar{\lambda}_{aG^0 G^\pm H^\pm} = 
    \frac{1}{2 K_0}
        \varepsilon^{abc} \bar{k}^b 
        (x^c + i y^c)
        (m_{H^\pm}^2 - m_c^2) 
\end{equation}

\noindent
{\bf $G^0$ - $G^0$ - $G^\pm$ - $H^\pm$}
\begin{equation}
    \bar{\lambda}_{G^0 G^0 G^\pm H^\pm} = 
    \frac{1}{2 K_0}
        \bar{k}^a 
        (x^a + i y^a)
        m_a^2 
\end{equation}

\subsection{Charged quartic couplings}

\noindent
{\bf $H^\pm$ - $H^\pm$ - $H^\pm$ - $H^\pm$}
\begin{equation}
    \bar{\lambda}_{H^\pm H^\pm H^\pm H^\pm} = 
    \frac{1}{K_0}
        \bigg(
        \bar{\tvec{k}}^\trans \bar{\mathcal M}^2_{\text neutral} \bar{\tvec{k}}
        -
        16 K_0 \bar{\tvec{\eta}}^\trans \bar{\tvec{k}}
        \bigg)
\end{equation}

\noindent
{\bf $G^\pm$ - $G^\pm$ - $H^\pm$ - $H^\pm$}
\begin{equation}
    \bar{\lambda}_{G^\pm G^\pm H^\pm H^\pm} = 
    \frac{1}{2 K_0}
        \bigg( 8 K_0
            \big(
            \eta_{00} - \frac{m_{H^\pm}^2}{4 K_0} + \bar{\tvec{\eta}}^\trans \bar{\tvec{k}}
            \big)
            - 2
             \bar{\tvec{k}}^\trans \bar{\mathcal M}^2_{\text neutral} \bar{\tvec{k}}
             +\tr \left(  \bar{\mathcal M}^2_{\text neutral} \right)
          \bigg)
\end{equation}

\noindent
{\bf $G^\pm$ - $G^\pm$ - $G^\pm$ - $G^\pm$}
\begin{equation}
    \bar{\lambda}_{G^\pm G^\pm G^\pm G^\pm} = 
    \frac{1}{K_0}
    \bar{\tvec{k}}^\trans \bar{\mathcal M}^2_{\text neutral} \bar{\tvec{k}}
\end{equation}

\noindent
{\bf $G^\pm$ - $H^\pm$ - $H^\pm$ - $H^\pm$}
\begin{equation}
    \bar{\lambda}_{G^\pm H^\pm H^\pm H^\pm} = 
        \big(
        \bar{\eta}^a - \bar{k}^a \frac{m_a^2}{K_0}
        \big)
        (x^a + i y^a)
\end{equation}

\noindent
{\bf $G^\pm$ - $G^\pm$ - $G^\pm$ - $H^\pm$}\noindent
\begin{equation}
    \bar{\lambda}_{G^\pm G^\pm G^\pm H^\pm} = 
    \frac{1}{K_0}
         \bar{k}^a 
        (x^a + i y^a)
        m_a^2
\end{equation}

\subsection{Vector - scalar couplings}
\label{sub:vectorscalar}
\noindent
{\bf $h_a$ - $G^0$ - $Z$}
\begin{equation}
    \bar{g}_{aG^0 Z} = 
    2 \frac{\sqrt{m_Z}}{\sqrt{K_0}} \bar{k}^a
\end{equation}

\noindent
{\bf $H^+$ - $H^+$ - $Z$}
\begin{equation}
    \bar{g}_{H^+ H^+ Z} = - i \sqrt{2}
    \frac{m_Z - 2 m_W}{\sqrt{K_0 m_Z}}
\end{equation}

\noindent
{\bf $G^+$ - $G^+$ - $Z$}
\begin{equation}
    \bar{g}_{G^+ G^+ Z} = - i \sqrt{2}
    \frac{m_Z - 2 m_W}{\sqrt{K_0 m_Z}}
\end{equation}

\noindent
{\bf $H^+$ - $H^+$ - $\gamma$}
\begin{equation}
    \bar{g}_{H^+ H^+ \gamma} = 2 i
    \frac{\sqrt{2 m_W (m_Z - m_W)}}{\sqrt{K_0 m_Z}}
\end{equation}

\noindent
{\bf $G^+$ - $G^+$ - $\gamma$}
\begin{equation}
    \bar{g}_{H^+ H^+ \gamma} = 2 i
    \frac{\sqrt{2 m_W (m_Z - m_W)}}{\sqrt{K_0 m_Z}}
\end{equation}

\noindent
{\bf $h_a$ - $H^+$ - $W^+$}
\begin{equation}
    \bar{g}_{a H^+ W^+} =  i \frac{\sqrt{2 m_W}}{\sqrt{K_0}}
    \left( x^a + i y^a \right)
 \end{equation}

\noindent
{\bf $h_a$ - $G^+$ - $W^+$}
\begin{equation}
    \bar{g}_{a G^+ W^+} =  2 \frac{\sqrt{m_W}}{\sqrt{2K_0}}
    \bar{k}^a
 \end{equation}

\noindent
{\bf $G^0$ - $G^+$ - $W^+$}
\begin{equation}
    \bar{g}_{G^0 G^+ W^+} =  -i 
    \frac{\sqrt{2m_W}}{\sqrt{K_0}}
 \end{equation}

\noindent
{\bf $h_a$ - $Z$ - $Z$}
\begin{equation}
    \bar{g}_{a Z Z} = 
    4 \frac{m_Z}{\sqrt{2K_0}} \bar{k}^a
\end{equation}

\noindent
{\bf $h_a$ - $W^+$ - $W^+$}
\begin{equation}
    \bar{g}_{a W^+ W^+} = 
    4 \frac{m_W}{\sqrt{2K_0}} \bar{k}^a
\end{equation}

\noindent
{\bf $G^+$ - $Z$ - $W^+$}
\begin{equation}
    \bar{g}_{G^+ Z  W^+} =  2 i \frac{\sqrt{2 m_W}}{\sqrt{K_0 m_Z}}
    (m_W-m_Z)
 \end{equation}

\noindent
{\bf $G^+$ - $\gamma$ - $W^+$}
\begin{equation}
    \bar{g}_{G^+ \gamma  W^+} =  2 i \frac{\sqrt{2(m_Z-m_W)}}{\sqrt{K_0 m_Z}}
    m_W
 \end{equation}

\noindent
{\bf $h_a$ - $h_b$ - $Z$ - $Z$}
\begin{equation}
    \bar{g}_{abZZ} = 
    \frac{4 m_Z^2}{K_0} \delta^{ab}
\end{equation}

\noindent
{\bf $h_a$ - $h_b$ - $W^+$ - $W^+$}
\begin{equation}
    \bar{g}_{abW^+W^+} = 
    \frac{4 m_W^2}{K_0} \delta^{ab}
\end{equation}

\noindent
{\bf $h_a$ - $H^+$ - $Z$ - $W^+$}
\begin{equation}
    \bar{g}_{aH^+ZW^+} =
        \frac{2\cos(\theta_W)}{K_0}
        \bigg(
        (2 m_W^2 - m_Z^2) (x^a - i y^a) 
        +
        i m_Z^2 \varepsilon^{abc} (x^b - i y^b) \bar{k}^b
        \bigg)
 \end{equation}

\noindent
{\bf $h_a$ - $G^+$ - $Z$ - $W^+$}
\begin{equation}
    g_{aG^+ZW^+} = 4 i 
    \frac{m_W}{m_Z}
    \frac{m_W^2 - m_Z^2}{K_0}
    \bar{k}^a
\end{equation}

\noindent
{\bf $h_a$ - $H^+$ - $\gamma$ - $W^+$}
\begin{equation}
    \bar{g}_{aG^+\gamma W^+} = 4 
    \frac{m_W^2}{K_0}
    \frac{\sqrt{m_Z^2-m_W^2}}{m_Z}
    (x^a - i y^a)
\end{equation}

\noindent
{\bf $h_a$ - $G^+$ - $\gamma$ - $W^+$}
\begin{equation}
    \bar{g}_{aG^+\gamma W^+} = 4 i
    \frac{m_W^2}{K_0}
    \frac{\sqrt{m_Z^2-m_W^2}}{m_Z}
    \bar{k}^a
\end{equation}

\noindent
{\bf $G^0$ - $G^0$ - $Z$ - $Z$}
\begin{equation}
    \bar{g}_{G^0 G^0 ZZ} = 4 \frac{m_Z^2}{K_0}
 \end{equation}

\noindent
{\bf $G^0$ - $G^0$ - $W^+$ - $W^+$}
\begin{equation}
    \bar{g}_{G^0 G^0 W^+W^+} = 4 \frac{m_W^2}{K_0}
 \end{equation}

\noindent
{\bf $G^0$ - $G^+$ - $Z$ - $W^+$}
\begin{equation}
    \bar{g}_{G^0G^+ Z W^+} = 4 
    \frac{m_W}{m_Z}
    \frac{m_Z^2-m_W^2}{K_0}
 \end{equation}

\noindent
{\bf $G^0$ - $G^+$ - $\gamma$ - $W^+$}
\begin{equation}
    \bar{g}_{G^0 G^+ \gamma W^+} = -4 
    \frac{m_W^2}{K_0} 
    \frac{\sqrt{m_Z^2-m_W^2}}{m_Z}
 \end{equation}

\noindent
{\bf $H^+$ - $H^+$ - $Z$ - $Z$}
\begin{equation}
    \bar{g}_{H^+ H^+ Z Z} = 4 
    \frac{(m_Z^2 - 2 m_W^2)^2}{K_0 m_Z^2}
 \end{equation}

\noindent
{\bf $G^+$ - $G^+$ - $Z$ - $Z$}
\begin{equation}
    \bar{g}_{G^+ G^+ Z Z} = 4 
    \frac{(m_Z^2 - 2 m_W^2)^2}{K_0 m_Z^2}
 \end{equation}

\noindent
{\bf $H^+$ - $H^+$ - $Z$ - $\gamma$}
\begin{equation}
    \bar{g}_{H^+ H^+ Z Z} = 8 m_W
    (2 m_W^2 - m_Z^2)
    \frac{\sqrt{m_Z^2 - m_W^2}}{K_0 m_Z^2}
 \end{equation}

\noindent
{\bf $G^+$ - $G^+$ - $Z$ - $\gamma$}
\begin{equation}
    \bar{g}_{G^+ G^+ Z \gamma} = 8 m_W
    (2 m_W^2 - m_Z^2)
    \frac{\sqrt{m_Z^2 - m_W^2}}{K_0 m_Z^2}
 \end{equation}

\noindent
{\bf $H^+$ - $H^+$ - $\gamma$ - $\gamma$}
\begin{equation}
    \bar{g}_{H^+ H^+ \gamma \gamma} = 8 m_W^2
    \frac{m_Z^2 - m_W^2}{K_0 m_Z^2}
 \end{equation}

\noindent
{\bf $G^+$ - $G^+$ - $\gamma$ - $\gamma$}
\begin{equation}
    \bar{g}_{G^+ G^+ \gamma \gamma} = 8 m_W^2
    \frac{m_Z^2 - m_W^2}{K_0 m_Z^2}
 \end{equation}

\noindent
{\bf $H^+$ - $H^+$ - $W^+$ - $W^+$}
\begin{equation}
    \bar{g}_{H^+ H^+ W^+ W^+} = \frac{4 m_W^2}{K_0}
\end{equation}

\noindent
{\bf $G^+$ - $G^+$ - $W^+$ - $W^+$}
\begin{equation}
    \bar{g}_{G^+ G^+ W^+ W^+} = \frac{4 m_W^2}{K_0}
\end{equation}

\subsection{Yukawa couplings}
\label{app:yukawaCouplings}

We note that we only give here the Yukawa couplings involving right-handed Weyl spinors. Their analogous parts for left-handed spinors are simply obtained through complex conjugation.\\

\noindent{\bf $ h $ - $f_L^\dagger$ - $f_R$}
\begin{subequations}
\begin{align}
    \bar{\xi}^u_a &= \frac{1}{\sqrt{2 K_0}} \, \left(\sigma_a\right)^\alpha_{\ \, \beta} \bar{\kappa}^*_\alpha \bar{\mathcal{U}}^\beta \label{eq:hupYuk}\\
    \bar{\xi}^d_a &= \frac{1}{\sqrt{2 K_0}} \, \left(\sigma_a\right)^\alpha_{\ \, \beta} \bar{\mathcal{D}}_\alpha \bar{\kappa}^\beta \label{eq:hdownYuk} \\
    \bar{\xi}^e_a &= \frac{1}{\sqrt{2 K_0}} \, \left(\sigma_a\right)^\alpha_{\ \, \beta} \bar{\mathcal{E}}_\alpha \bar{\kappa}^\beta
\end{align}
\end{subequations}

\noindent{\bf $G^0$ - $f_L^\dagger$ - $f_R$}
\begin{subequations}
\begin{align}
    \bar{\xi}^u_{G^0} &= \frac{i}{\sqrt{2 K_0}} \, \bar{\kappa}^*_\alpha \bar{\mathcal{U}}^\alpha\\
    \bar{\xi}^d_{G^0} &= -\frac{i}{\sqrt{2 K_0}} \, \bar{\mathcal{D}}_\alpha \bar{\kappa}^\alpha\\
    \bar{\xi}^e_{G^0} &= -\frac{i}{\sqrt{2 K_0}} \, \bar{\mathcal{E}}_\alpha \bar{\kappa}^\alpha
\end{align}
\end{subequations}

\noindent{\bf $H^\pm$ - $f_L^\dagger$ - $f'_R$}

\begin{subequations}
\begin{align}
    \bar{\xi}^{ud}_{H^+} &= -\frac{i}{\sqrt{K_0}} \varepsilon_{\alpha\beta} \bar{\kappa}^\alpha \bar{\mathcal{U}}^\beta\\
    \bar{\xi}^{ud}_{H^-} &=  -\frac{i}{\sqrt{K_0}} \varepsilon^{\alpha\beta} \bar{\mathcal{D}}_\alpha \bar{\kappa}^*_\beta\\
    \bar{\xi}^{\nu e}_{H^-} &= -\frac{i}{\sqrt{K_0}} \varepsilon^{\alpha\beta} \bar{\mathcal{E}}_\alpha \bar{\kappa}^*_\beta
\end{align}
\end{subequations}

\noindent{\bf $ G^\pm - f_L^\dagger - f'_R$}

\begin{subequations}
\begin{align}
    \bar{\xi}^{ud}_{G^+} &= \frac{i}{\sqrt{K_0}} \bar{\kappa}^*_\alpha \bar{\mathcal{U}}^\alpha\\
    \bar{\xi}^{ud}_{G^-} &=  \frac{i}{\sqrt{K_0}} \bar{\mathcal{D}}_\alpha \bar{\kappa}^\alpha\\
    \bar{\xi}^{\nu e}_{G^-} &= \frac{i}{\sqrt{K_0}} \bar{\mathcal{E}}_\alpha \bar{\kappa}^\alpha
\end{align}
\end{subequations}

\acknowledgments

This work is supported by Campus France/DAAD, project PROCOPE 46704WF, by {\it Investissements d’avenir}, Labex ENIGMASS, contrat ANR-11-LABX-0012, by Chile ANID FONDECYT project 1200641, and by the IN2P3 master project “Th\'eorie – BSMGA”.
We are grateful to the Centre de Physique Th\'eorique Grenoble Alpes (CPTGA) for the hospitality extended to M.~Maniatis during his stay in Grenoble in fall 2021.

\bibliographystyle{JHEP}
\bibliography{references}

\end{document}